\documentclass[twocolumn]{aastex631}

\usepackage{subfigure}
\usepackage{color}
\usepackage{mathrsfs}
\usepackage{amsmath}
\usepackage{cases}    
\usepackage{lmodern}

\begin{document}

\title{Study for curvature radiation and magnetic pair creation process on polar-cap region of magnetic white dwarf}

\author{Bingyan Wang}
\email{d202380099@hust.edu.cn}
\affiliation{Department of Astronomy, School of Physics, Huazhong 
     University of Science and Technology, Wuhan, Hubei 430074, China}

\author{Jumpei Takata}

\affiliation{Department of Astronomy, School of Physics, Huazhong 
     University of Science and Technology, Wuhan, Hubei 430074, China}

\email{ takata@hust.edu.cn}




\begin{abstract}
 Rapidly rotating, strongly magnetized white dwarfs (WDs) have been proposed as potential sites of rotation-powered activity analogous to that of a neutron star pulsar. In this study, we investigate particle acceleration, radiation processes, pair creation and resulting synchrotron radiation in the polar cap acceleration region. 
 Within the framework of the space-charge-limited flow model, we examine how these processes depend on the spin period and surface magnetic field using both one-dimensional numerical calculations and analytical estimates.  
 To explore the impact of the magnetic field geometry on the accelerating process, we consider  both a pure dipole field and a combination of dipole and quadrupole fields.  
 The inclusion of a quadrupole component reduces the curvature radius of the magnetic field lines, and significantly enhances the accelerating field, leading to more efficient radiation and pair creation processes. Using this framework, we evaluate the WD death line with a more consistent treatment of the relevant physical processes than the previous studies.
 We find that the pair creation process can occur for spin periods $P\lesssim100$~s, when the dipole field strength $B_{*,d}\lesssim10^{10}$~G, indicating that pair creation is difficult to sustain in currently known magnetic WDs. We discuss the implications of our model for rotation-powered activity in rapidly spinning isolated magnetic WDs and for the possible WD interpretation of long-period radio transients.
\end{abstract}

\keywords{
acceleration of particles ---
radiation mechanisms: non-thermal ---
white dwarfs ---
magnetic fields ---
pulsars: general
}


\section{Introduction} \label{sec:intro}

Pulsars have been recognized as rapidly spinning and highly magnetized neutron stars (hereafter NS), and their radio emission is generally attributed to a coherent radiation process, exhibiting a high brightness temperature of the order of $\sim 10^{22} \mathrm{K}$ ~\citep[for example, figure~1 of ][]{2018NatAs...2..865K}. It has been widely accepted that the region of radio emission originates from the polar cap region near the magnetic axis. The polar cap is defined as the surface region from which the global dipolar magnetic field lines extend to the light cylinder, which is the position where the co-rotation speed with the NS becomes the speed of light. The axial distance of the light cylinder from the stellar spin axis is defined by $R_{lc}\equiv cP/(2\pi)$ with $P$ denoting the stellar spin period.

It has been argued that an electric field parallel to the magnetic field lines can develop above the polar cap if the local charge density deviates from the so-called Goldreich-Julian (hereafter GJ) charge density~\citep{1969ApJ...157..869G}, thereby accelerating electrons (or possibly positrons or protons) extracted from the neutron star (NS) surface \citep{1971ApJ...164..529S,1975ApJ...196...51R,2002nsps.conf..240U,2001ApJ...560..871H,2002ApJ...565..482G}. The accelerated particles emit high-energy gamma rays through curvature radiation~\citep{1975ApJ...196...51R} or inverse Compton scattering~\citep{1996A&A...310..135Z,2002ApJ...568..862H}. The emitted gamma rays are converted into electron--positron pairs through magnetic pair creation. These newly created pairs can emit synchrotron photons, which subsequently trigger a pair creation cascade. The pair multiplicity, denoted by $\kappa$ in this study, is defined as the number of electron--positron pairs produced per primary particle accelerated in the polar-cap accelerator. \citet{2015ApJ...810..144T} argued that the maximum pair multiplicity in pulsar magnetospheres is of the order of $\kappa \sim \mathrm{few} \times 10^5$.

As the pulsar spins down, the pair multiplicity decreases because the accelerating electric field becomes weaker. The so-called death line of pair creation, or equivalently the cessation of pulsar activity on the $P$--$\dot{P}$ diagram, has been predicted by different models~\citep{1993ApJ...402..264C,2022MNRAS.510.2572B,2022MNRAS.516.5084B}. Recent deep radio observations have discovered slow-rotating pulsars located near the death line on the $P$--$\dot{P}$ diagram, such as PSR J0901$-$4046, with a spin period of $P \sim 23.5~{\rm s}$ and a period derivative of $\dot{P} \sim 2.7 \times 10^{-14}~{\rm s\,s^{-1}}$~\citep{2018ApJ...866...54T}, and PSR J0250+5854, with $P \sim 76~{\rm s}$ and $\dot{P} \sim 2.25 \times 10^{-13}~{\rm s\,s^{-1}}$~\citep{2022NatAs...6..828C}.

Beyond NS pulsars, the polar-cap acceleration mechanism has also been explored in the context of rapidly rotating, strongly magnetized white dwarfs (WDs). In this paper, we revisit the rotation-powered activity of highly magnetized WDs, with a particular focus on the radiation and pair creation processes occurring near the polar cap. Our motivation stems from the recent discoveries of (i) highly magnetized and rapidly spinning {\it isolated} WDs exhibiting X-ray emission, such as ZTF J2008+4449~\citep{2026A&A...706A.188C} and J1901+1458~\citep{2021Natur.595...39C,2024PASJ...76..702B,2025arXiv250903216D}; (ii) non-thermal emission from WD/M-dwarf binary systems, including AR~Scorpii~\citep{2016Natur.537..374M,2018ApJ...853..106T} and J191213.72$-$441045.1~\citep{2023NatAs...7..931P}; and (iii) so-called long-period radio transients (hereafter LPTs). LPTs are a newly identified class of coherent periodic radio sources whose observed periods are significantly longer than those of typical pulsars~\citep{2022Natur.601..526H,2023Natur.619..487H,2026arXiv260110393R}.
Among the LPTs, CHIME J0630+25 is known as the shortest-period source, with a period of $\sim 421$~s~\citep{2024arXiv240707480D}, while ASKAP J1839$-$0756 is the longest-period source, with a period of $P = 387$~minutes~\citep{2025NatAs...9..393L}. If LPTs are indeed conventional young NSs with a typical spin-down magnetic field of $B \sim 10^{12}$~G, they would lie below the death line in the $P$--$\dot{P}$ (or $P$--$B$) diagram. Although the spin-down rates of LPTs have not yet been measured precisely, their long periods suggest that they likely represent a class of radio sources distinct from conventional pulsars. Hence, it has been theoretically investigated
  that LPTs are related to the rotation powered activities
  of the rapidly spinning WDs~\citep{Katz2022,2024ApJ...961..214R,2026JHEAp..5300593M}. Intriguingly, two LPTs, ILT~J1101+5521 with $P \sim 125.5$~minutes and GLEAM-X J0704$-$37 with $P \sim 174.9$~minutes, have been identified as WD--M-dwarf binaries, and the observed periods have been interpreted as either the spin periods of the WDs or the orbital periods of the binary systems~\citep{2025NatAs...9..672D,2025A&A...695L...8R}.  Recently, new LPT, ASKAP J174508.9-505149, is classified as an accreting cataclysmic variable~\citep{2026arXiv260604232R}. These observations suggest that LPTs may include both WD-related systems and neutron-star systems.

If we apply the NS-like rotation-powered scenario to WDs, the available electric potential induced by the WD spin becomes of the order of
\begin{equation}
\Phi_a=(2\pi)^2\frac{\mu_{WD}}{2c^2P^2}\sim 9\times 10^9\left(\frac{\mu_{WD}}{10^{35}{\rm G\,cm^3}}\right)
\left(\frac{P}{500~{\rm s}}\right)^{-2}~{\rm statvolt},
\label{avap}
\end{equation}
where $\mu_{WD}$ is the dipole moment of the WD. Here, the physical quantities are scaled to the characteristic values of ZTF J2008+4449 and J1901+1458, which have spin periods of $P\sim 390$~s and $\sim 416$~s, respectively, and surface magnetic fields of the order of $B\sim (0.5$--$1)\times 10^9$~G. This available potential can, in principle, accelerate charged particles to relativistic energies, with a maximum Lorentz factor of $\gamma_{\max}\equiv e\Phi_a/m_ec^2\sim 5\times 10^6(\mu_{WD}/10^{35}~{\rm G\,cm^3})(P/500~{\rm s})^{-2}$. The corresponding spin-down power due to magnetic dipole radiation is
\begin{equation}
  L_{sd}=\frac{(2\pi)^4\mu_{WD}^2}{6c^3P^4}\sim 2\times 10^{30}~{\rm erg~s^{-1}}
  \left(\frac{\mu_{WD}}{10^{35}~{\rm G\,cm^3}}\right)^2\left(\frac{P}{500~{\rm s}}\right)^{-4}.
\end{equation}
These estimates suggest that rapidly spinning, highly magnetized WDs may produce detectable radiation, provided that the rotational energy loss is efficiently converted into radiation. However, compared to NSs, the curvature radiation process may be  strongly suppressed because the curvature radius of the magnetic field lines in the WD magnetosphere is much larger. To investigate the possibility of rotation-powered activity in the aforementioned WDs and LPTs, it is therefore important to examine the radiation and pair creation processes in detail.

Although definitive observational evidence is still lacking, high-energy X-ray and gamma ray emission driven by rotation-powered activity in WDs has been explored in previous studies~\citep{1988SvAL...14..258U,1998A&A...338..521I,2011PhRvD..83b3002K,2017ApJ...851..143T}. For instance, \citet{1998A&A...338..521I} and \citet{2017ApJ...851..143T} discussed curvature radiation and inverse Compton scattering in WD magnetospheres and evaluated the possible high-energy emission from WD binary systems such as AE~Aquarii and AR~Scorpii. \citet{2011PhRvD..83b3002K} investigated pair creation in the WD magnetosphere as a possible source of cosmic-ray electrons and positrons. Furthermore, \citet{2011PhRvD..83b3002K} and \citet{2012JPhG...39f5001B} evaluated the death line for magnetic pair creation in WD pulsars. While these studies discussed particle acceleration and the resultant high-energy emission in WD magnetospheres, a self-consistent quantitative investigation that combines the accelerating electric field, curvature radiation, and the resulting magnetic pair creation cascade remains lacking. In particular, previous studies have not evaluated the pair multiplicity in WD magnetospheres. In this paper, we investigate whether WDs can sustain a pair creation cascade above their polar-cap regions and predict the corresponding pair multiplicity. We also explore the implications for rapidly spinning, highly magnetized WDs. The theoretical framework is described in Section~\ref{sec:model}, and the resulting pair multiplicities and death line are presented in Section~\ref{sec:result}. A discussion and summary of our results are given in Sections~\ref{sec:discussion} and~\ref{summary}, respectively.

\section{Theoretical model}
\label{sec:model}
In this section, we describe our theoretical model for particle acceleration and electron--positron pair creation above the polar cap of WDs. The model consists of three main components: (i) particle acceleration and curvature radiation of electrons extracted from the stellar surface, (ii) magnetic pair creation, and (iii) synchrotron radiation from secondary pairs. A key assumption and simplification in this study concerns the geometry of the acceleration and pair creation region. Specifically, we adopt a one-dimensional model along a magnetic field line to evaluate the pair creation cascade, following the method developed by \citet{2015ApJ...810..144T} for NS pulsars. Strictly speaking, this one-dimensional treatment is valid only when high-energy photons do not travel significantly across the magnetic field lines before converting into pairs. In canonical NS pulsars, this assumption is expected to hold because the mean free path (hereafter MFP) for magnetic pair creation of gamma rays emitted near the stellar surface can be much shorter than the polar-cap size. In contrast, even for strongly magnetized WDs with a surface magnetic field of $B \sim 10^{8-9}$~G, the MFP and the characteristic size of the acceleration region are typically larger than the polar-cap size (Section~\ref{sec:photon_absorption}), indicating that transverse motion relative to the magnetic field lines may not be negligible. Nevertheless, in this work, we adopt the one-dimensional model as a first step toward evaluating the characteristics of pair creation in WD magnetospheres.  In section~\ref{threeg}, we discuss the influence of the three-dimensional geometry on the pair-creation multiplicity.

\subsection{Particle acceleration in polar cap}
We apply a particle-acceleration model analogous to that used for NS pulsars, in which an electric-field component parallel to the magnetic field lines is induced near the polar cap. The size of the polar cap of WDs is characterized by
\begin{eqnarray}
  R_{pc}&=&R_{WD}\left(\frac{R_{WD}}{R_{lc}}\right)^{1/2} \nonumber \\
    &\sim&7\times 10^6\left(\frac{R_{WD}}{5\cdot 10^8~{\rm cm}}\right)^{3/2}
    \left(\frac{P}{500~{\rm s}}\right)^{-1/2}~{\rm cm}.
\label{rpc}
\end{eqnarray}
The so-called non-corotational potential, which gives rise to the electric field parallel to the magnetic field, is described by the Poisson equation,
\begin{equation}
  \nabla^2\Phi_{nco}=-4\pi (\rho-\rho_{GJ}),
  \label{poisson}
\end{equation}
where $\rho_{GJ}=-\boldsymbol{B}\cdot\boldsymbol{\Omega}/2\pi c$ is the GJ charge density, with $\boldsymbol{B}$ being the local magnetic field and $\boldsymbol{\Omega}$ the angular velocity vector of the star. The electric field parallel to the magnetic field is calculated from $E_{||}=-\partial \Phi_{nco}/\partial s$, where $s$ is the distance measured along the magnetic field line.

Extensive research on the polar-cap accelerator in NSs has been carried out~\citep[e.g.,][]{1975ApJ...196...51R,1978ApJ...222..297S,1983ApJ...266..215A,1997ApJ...485..735M,2024ApJ...974L..32C}. The expression $E_{||}=2\Omega B_s s/c$~\citep{1975ApJ...196...51R} is often used to estimate the characteristic accelerating electric field. This expression is valid under two key conditions: (i) the time-averaged density of the accelerating electrons is much lower than the GJ value, and (ii) the radial distance to the magnetic pair creation front, where the electric field begins to be screened, is much smaller than the polar-cap size defined by equation~(\ref{rpc}). When the accelerator extends radially beyond the polar-cap size, the accelerating electric field may be better described by the so-called space-charge-limited flow (SCLF) model~\citep{1978ApJ...222..297S,1997ApJ...485..735M}. In this framework, electrons are freely extracted from the stellar surface and flow along the magnetic field lines. The local charge density of this flow is proportional to $B$, whereas the local GJ charge density is proportional to the component $B_z$ along the spin axis. The difference between the actual and GJ charge densities develops along the curved magnetic field lines and leads to the formation of an accelerating electric field. As we discuss in Section~\ref{sec:photon_absorption}, the MFP for magnetic pair creation is longer than the polar-cap size for the typical surface magnetic field $B_s=10^{8-9}$~G of magnetized WDs. In this study, therefore, we apply the SCLF model to investigate the polar-cap accelerator in WDs.

\subsection{Acceleration with dipole field}
\label{sec:dipole_field_intro}
For the SCLF model, the {\it local} magnetic-field geometry, through its effect on $B_z$, is a crucial factor in determining the accelerating electric field. In this work, therefore, we examine polar-cap acceleration under two magnetic-field configurations: (i) a pure dipole field and (ii) a combination of dipole and quadrupole fields (Section~\ref{nondipole}). As demonstrated by \citet{1982ApJ...254..713B}, the combination of dipole and quadrupole fields can produce stronger acceleration near the stellar surface. In the NS case, general-relativistic effects influence the magnitude of the accelerating electric field~\citep{1997ApJ...485..735M}, whereas such effects are negligible for WDs.

For the pure-dipole case, we apply the analytical expression for the non-corotational potential obtained by \citet{1978ApJ...222..297S}:
\begin{eqnarray}
    \Phi_{nco}(r,\theta_*,\phi_*)&=&\frac{3}{4}
    \Phi_a\theta_c
    \left(\frac{\theta_*}{\theta_c}\right)
    \left( 1- \frac{\theta_* ^2}{\theta_c^2}\right) \nonumber \\
&\times&
    \left[  \left( \frac{r}{R_{\text{WD}}}\right)^{1/2} -1 \right]
  \cos \phi_* \, ,
\label{phimax}
\end{eqnarray}
where $\theta_c=(R_{WD}/R_{lc})^{1/2}$ is the opening angle of the polar field lines at the stellar surface, and $\theta_*$ and $\phi_*$ are the magnetic colatitude and azimuth, respectively, measured about the dipole axis. Strictly speaking, the expression in equation~(\ref{phimax}) is valid for an orthogonal rotator, in which the angle between the spin and magnetic axes is $90^\circ$. Nevertheless, we adopt this potential drop because the orthogonal rotator yields the maximum potential drop near the stellar surface owing to the more rapid variation of $B_z$ along the magnetic field lines. This allows us to place tight constraints on the spin period ($P$) and surface dipole field ($B_{*,d}$) required for pair creation to occur in WDs.

Within the one-dimensional framework, we evaluate the potential drop at $\theta_*=\theta_c/2$ and $\phi_*=0$,
 \begin{equation}
   \Phi_{nco}(r)=\frac{9}{16}\Phi_a\theta_c\left[ \left( \frac{r}{R_{\text{WD}}}\right)^{1/2} -1 \right].
   \label{phinco}
 \end{equation}
Assuming $r\sim s+R_{WD}$, the electric field parallel to the magnetic field is estimated as
\begin{equation}
  E_{||}(s)=-\frac{d\Phi_{nco}}{ds}=-\frac{9}{32}\frac{\Phi_a}{(rR_{WD})^{1/2}}\theta_c.
\label{epara1}
\end{equation}
Taking into account the radiation-reaction force due to curvature radiation, the evolution of the Lorentz factor, $\gamma$, is calculated from
\begin{equation}
     m_ec^2\frac{d\gamma}{dt}=-ceE_{||}-P_c,
\label{eq:gam}
\end{equation}
where $P_c=2e^2c\gamma^4/(3\rho_c^2)$ is the curvature-radiation power, and $\rho_c$ is the curvature radius of the magnetic field line.

Since the strength of a dipole magnetic field decreases as $B\propto r^{-3}$, magnetic pair creation in the WD magnetosphere, if it occurs, is expected to take place near the stellar surface. In this study, therefore, we evaluate the pair creation process within $r<2R_{WD}$. The Lorentz factor attainable through the potential drop within $R_{WD}<r<2R_{WD}$ is
\begin{eqnarray}
  \gamma &=& \frac{e\Phi_{nco}(2R_{WD})}{m_ec^2} \nonumber \\
  &\sim& 2.3\times 10^4\left(\frac{\mu_{WD}}{10^{35}~{\rm G\,cm^3}}\right)
  \left(\frac{P}{500~{\rm s}}\right)^{-3/2}\left(\frac{R_{WD}}{10^9~{\rm cm}}\right)^{1/2}.
\label{gamma}
\end{eqnarray}
The corresponding typical energy of curvature photons is
\begin{eqnarray}
  E_c&=&\frac{3}{2}\frac{\hbar c}{\rho_c}\gamma^3 
  \sim 0.4\left(\frac{\rho_c}{10^9~{\rm cm}}\right)^{-1}
  \left(\frac{\mu_{WD}}{10^{35}~{\rm G\,cm^3}}\right)^3 \nonumber \\
  &\times& \left(\frac{P}{500~{\rm s}}\right)^{-9/2}\left(\frac{R_{WD}}{10^9~{\rm cm}}\right)^{3/2}~{\rm eV}.
  \label{ec}
\end{eqnarray}
We find that, within the pure-dipole field configuration, the characteristic energy of curvature photons remains below the electron rest-mass energy, which prevents magnetic pair creation. We therefore conclude that pair creation becomes feasible only for WDs that are both extremely rapidly rotating ($P<100$~s) and strongly magnetized ($\mu_{WD}>10^{35}~{\rm G\,cm^{3}}$).

From equation~(\ref{phinco}), we find that the potential drop near the stellar surface is suppressed by a factor of $\theta_c \sim 0.02 \left(P/500~{\rm s}\right)^{-1/2}\left(R_{WD}/10^{9}~{\rm cm}\right)^{1/2}$ compared to the available potential drop given in equation~(\ref{avap}). This reduction is related to the curvature of the magnetic field lines near the stellar surface and indicates how rapidly the charge density of the electron flow deviates from the GJ charge density along the magnetic field lines. Near the polar cap of a dipolar field, the curvature radius scales as $\rho_c \sim (R_{WD}R_{lc})^{1/2} \sim R_{WD}/\theta_c$ (up to a factor of order unity), so that $\Phi_{nco}$ near the surface is reduced relative to $\Phi_a$ by a factor of $\sim R_{WD}/\rho_c \sim \theta_c$. If the magnetic field lines have a larger curvature (i.e., a smaller curvature radius) than in the dipole configuration, the potential drop can exceed that of the dipole case described by equation~(\ref{phinco}), as investigated by \citet{1982ApJ...254..713B} for NS pulsars. Such an enhancement of the potential drop would lead to a higher Lorentz factor of the accelerated electrons and, consequently, to more energetic photons emitted near the stellar surface.

\begin{figure*}
    \centering

    \subfigure[$\zeta=0.5\,\mathrm{rad}$.]
    {
        \includegraphics[width=0.405\textwidth]{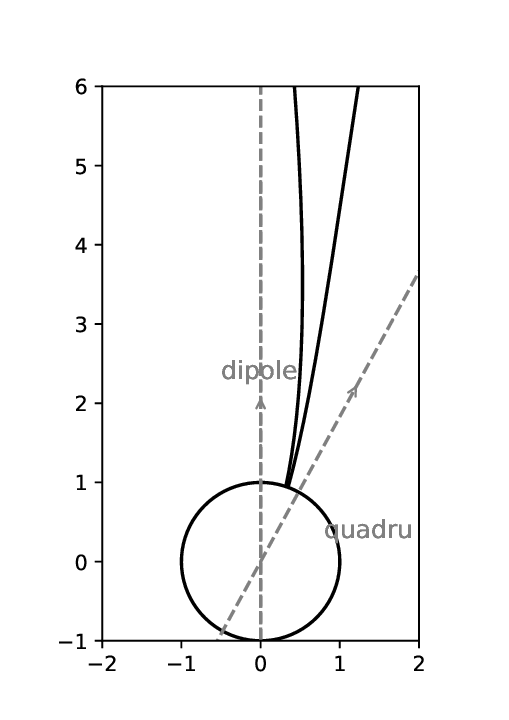}
        \label{fig:pc_boundary_a}
    }
    \hspace{0.02\textwidth}
    \subfigure[$\zeta=1.4\,\mathrm{rad}$.]
    {
        \includegraphics[width=0.38\textwidth]{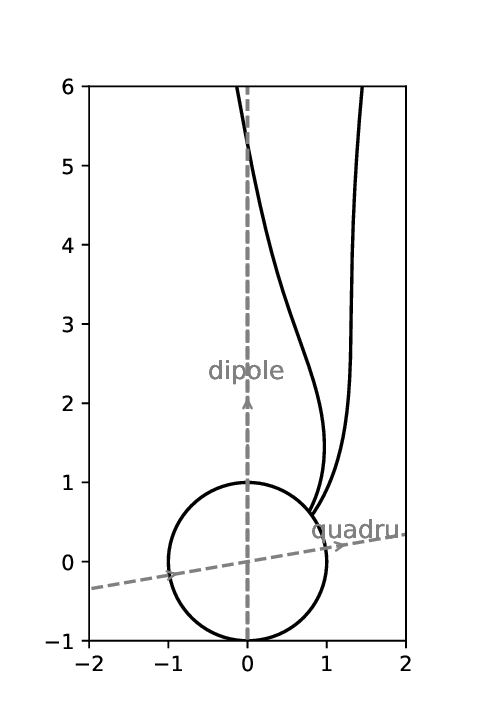}
        \label{fig:pc_boundary_b}
    }

    \caption{
    Configuration of the last-open field lines (solid curves) near the stellar surface for a magnetic field composed of dipole and quadrupole components. The dashed lines indicate the dipole and quadrupole axes, and the angles between the two axes are $\zeta=0.5$~rad (left panel) and $1.4$~rad (right panel). The results are shown for $B_{*,q}/B_{*,d} = 4$, $P=100~{\rm s}$, and $R_{WD}=5\times 10^8$~cm. The circles represent the WD, and the spatial scale is normalized to the stellar radius.}
    \label{fig:pc_boundary}
\end{figure*}

\subsection{Combination of dipole and quadrupole fields}
\label{nondipole}

\begin{figure*}
    \centering
    \subfigure{
    \raisebox{-0.1cm}{\includegraphics[width=0.44\textwidth]{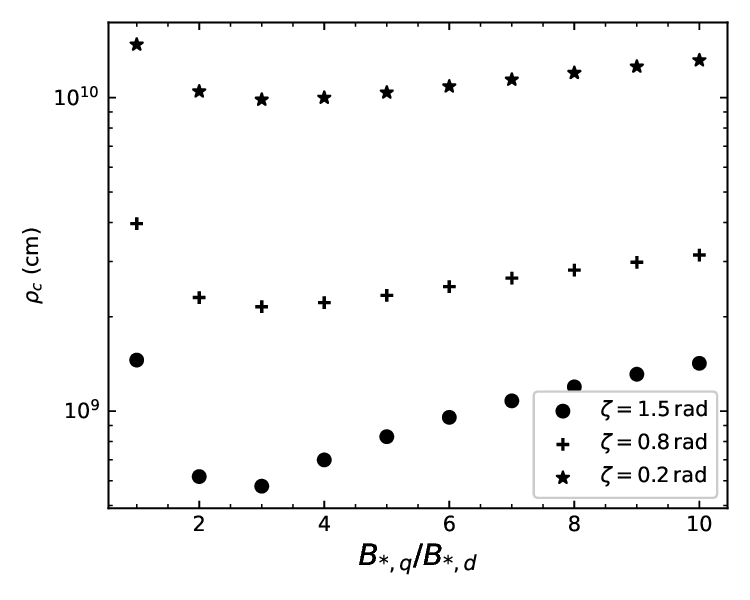}}
    }
    \quad
    \subfigure{
    \includegraphics[width=0.50\textwidth]{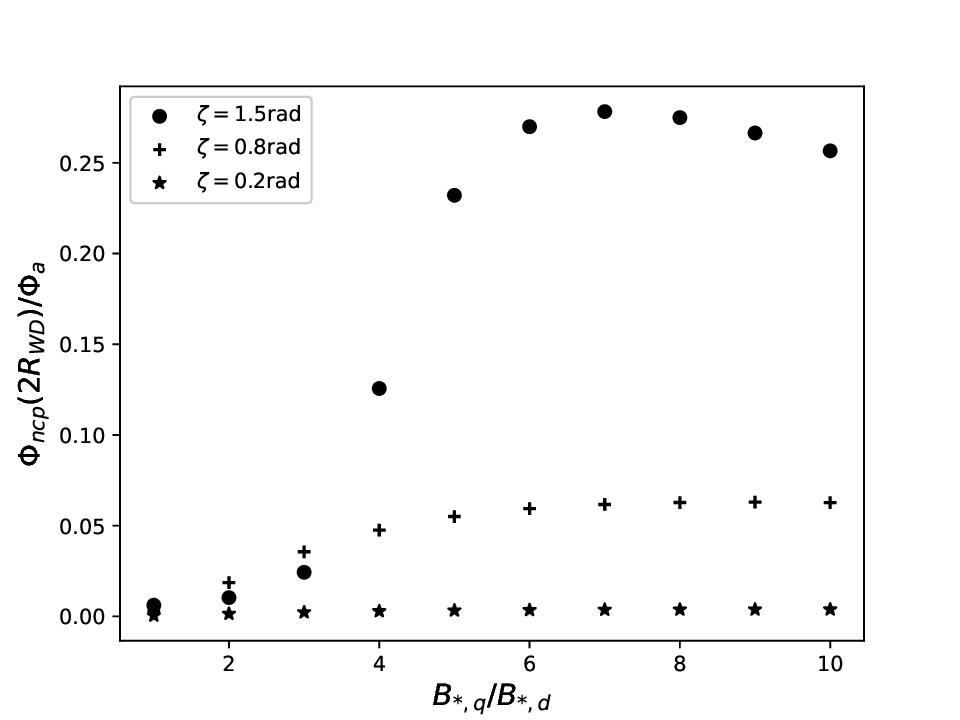}
    }
    \caption{Left: Curvature radius near the WD surface as a function of the field-strength ratio $\beta=B_{*,q}/B_{*,d}$.
      Right: The non-corotational potential $\Phi_{nco}\propto \Phi_a\left(\frac{B_z}{B}-\frac{B_{z,*}}{B_*}\right)$ at $r=2R_{WD}$. The results are shown for different inclination angles $\zeta$ between the dipole and quadrupole axes and correspond to a field line at the center of the magnetic flux tube. The spin period is assumed to be $P=500$~s.}
    \label{pratio}
\end{figure*}

Observational evidence suggests that, in some WDs, the magnetic field near the stellar surface includes non-dipolar components, such as quadrupole and/or octupole fields, which can even dominate over the dipole field~\citep[e.g.,][]{1984MNRAS.206..407M,2003ApJ...593.1040V,2005A&A...442..651E}. 
  For example,  \cite{2008ApJ...683..466V} carried out a Zeeman-tomographic analysis of the magnetic WD 1953-011 and obtained a  best-fit configuration with $B_{*,q}/B_{*,d}\sim1.3$, where $B_{*,q}$ is the quadrupole component. Similarly, the tomography analysis of HE~1045$-$0908~\citep{2005A&A...442..651E} revealed a
  quadrupole dominated field structure
  with $B_{*,q}/B_{*,d}\sim {\rm few}$. These results would suggest that  magnetic configurations in which
  the quadrupole component is comparable to or stronger than the dipole component may be
  relatively common among magnetic WDs.

  In addition, several highly magnetized and rapidly rotating massive WDs, such as ZTF J2008+4449 and J1901+1458, have been proposed as products of double-WD mergers. During the merger process, the magnetic field can be substantially amplified through
  dynamo mechanisms driven by differential rotation and magnetorotational instability in the merger remnant and
  its surrounding disk~\citep[e.g.][]{GarciaBerro2012, 2013ApJ...773..136J,2015ApJ...806L...1Z,2024A&A...691A.179P}. Such turbulent magnetic-field amplification is expected to generate complex magnetic topologies and may naturally give rise to non-dipolar surface field structures
  \citep{2026JHEAp..5300593M}. Motivated by these observational and theoretical indications, we investigate how the presence of a quadrupole component modifies the curvature of the magnetic field and its influence on the polar cap physical processes.

  Following \citet{1982ApJ...254..713B}, we evaluate the physical processes in the plane defined by the dipole and quadrupole axes and  express the dipole and quadrupole field configurations as
\begin{equation}
\boldsymbol{B}_{d}(r,\theta_d)
= B_{*,d}\left(\frac{R_{WD}}{r}\right)^{3}
\left( 2\cos\theta_d\,\hat{\boldsymbol{r}}
      + \sin\theta_d\,\hat{\boldsymbol{\theta}}_d \right)
\end{equation}
and
\begin{eqnarray}
\boldsymbol{B}_{q}(r,\theta_q)&=&B_{*,q}\left(\frac{R_{WD}}{r}\right)^{4} \nonumber \\
&\times&\left[ (3\cos^{2}\theta_q - 1)\,\hat{\boldsymbol{r}}
  + 2\cos\theta_q\sin\theta_q\hat{\boldsymbol{\theta}}_q \, \right],
\end{eqnarray}
respectively, where $\theta_d$ and $\theta_q$ are the angles measured from the dipole and quadrupole axes, respectively. In addition, $\hat{\boldsymbol{r}}$ and $\hat{\boldsymbol{\theta}}_{d}$ (or $\hat{\boldsymbol{\theta}}_{q}$) represent the unit vectors of the polar coordinates defined with respect to the dipole (or quadrupole) axis, respectively. The total magnetic field is given by
$\boldsymbol{B}(r,\theta,\phi)=\boldsymbol{B}_{d}(r,\theta_d)+\boldsymbol{B}_q(r,\theta_q)$. In this study, for simplicity, we assume that the dipole axis is aligned with the spin axis, which yields $\theta=\theta_d$. We then introduce $\beta\equiv B_{*,q}/B_{*,d}$ as the ratio of quadrupole to dipole field strengths at the surface, and $\zeta$ as the inclination angle between the dipole and quadrupole axes. Figure~\ref{fig:pc_boundary} shows the last-open field lines tangent to the light cylinder for $\beta=4$, with $\zeta=0.5$~rad in the left panel and $\zeta=1.4$~rad in the right panel. As shown in the figure, as $\zeta$ approaches $90^\circ$, the curvature radius becomes comparable to the stellar radius.  In this paper, we parameterize the strength ratio $\beta$ in a range of $0<\beta<10$, consistent with the observational constraints discussed above.

To estimate the non-corotational electric potential, we adopt the expression obtained by \citet{1982ApJ...254..713B},
\begin{eqnarray}
  \Phi_{nco}&\sim& 2\Phi_a\left(\frac{B_z}{B}-\frac{B_{z,*}}{B_*}\right)\frac{\Delta y/\Delta x}{1+(\Delta y/\Delta x)^2} \nonumber \\
  &\times&\left[1-\left(\frac{x}{\Delta x}\right)^2-\left(\frac{y}{\Delta y}\right)^2\right],
\label{q+d phi}
\end{eqnarray}
where $\Delta x$ and $\Delta y$ denote the semi-axes of the elliptical cross-section of the magnetic flux tube, while $x$ and $y$ represent the corresponding coordinates measured along those axes. As shown by \citet{1982ApJ...254..713B} and equation~(\ref{q+d phi}), the available potential drop (see equation~(\ref{avap})) in the pure-dipole case still represents the total potential drop across the accelerator $(\Phi_{nco}\propto \Phi_a)$. This is because the dipole component dominates the magnetic field near the light cylinder, and magnetic dipole radiation is the primary mechanism for the stellar spin-down. In our one-dimensional treatment, we evaluate the potential drop at the center of the flux tube ($x=y=0$) and estimate $\Delta x$ and $\Delta y$ in the plane defined by the dipole and quadrupole axes and in the perpendicular direction, respectively. The ratio $\Delta x/\Delta y$ is of order unity along the flux tube.

The SCLF model predicts a potential drop of the order of $\Phi_{nco} \sim \Phi_a (B_z/B-B_{z,*}/B_*)$, as shown in equation~(\ref{q+d phi}). From the right panel of Figure~\ref{fig:pc_boundary}, we see that the ratio $B_z/B$ for the combination of dipole and quadrupole fields can deviate substantially from the surface value $B_{*,z}/B_*$ within $r<2R_{WD}$. This variation significantly enhances the potential drop relative to the pure-dipole configuration.

The left and right panels of Figure~\ref{pratio} illustrate how the curvature radius $\rho_c$ and the potential drop $\Phi_{nco}(2R_{WD})/\Phi_a$, respectively, at $r=2R_{WD}$ depend on the inclination angle $\zeta$ and the field-strength ratio $\beta=B_{*,q}/B_{*,d}$; the triangles, crosses, and circles correspond to inclination angles of $\zeta=0.2$, 0.8, and 1.5~rad, respectively. The results are calculated for a period of $P=500$~s, although the dependence on $P$ is weak. As the left panel shows, a quadrupole component with a larger inclination angle significantly reduces the curvature radius of the field lines in the polar-cap region ($\rho_c\sim 3\times 10^{10}~{\rm cm}$ for the pure-dipole configuration). Accordingly, the potential drop near the stellar surface is enhanced. In particular, for the nearly perpendicular case with $\zeta=1.5$~rad, the potential drop can reach about 25\% of the available potential drop when the field-strength ratio reaches $B_{*,q}/B_{*,d}>5$. Higher-order multipole components may increase the potential drop even further. Therefore, when non-dipolar components dominate near the surface, a substantial fraction of the available potential drop can be realized within $r<2R_{WD}$.

If the radiation-reaction force due to curvature radiation remains negligible compared to the acceleration term in equation~(\ref{eq:gam}), electrons can be accelerated to 
$\gamma_{\max}\sim 1.5\times 10^6(f_r/0.3)(\mu_{WD}/10^{35}~{\rm G\,cm^3})(P/500~{\rm s})^{-2}$,
where $f_r\equiv \Phi_{nco}/\Phi_a$. On the other hand, when the drag force, which is proportional to $\gamma^4$, becomes significant, it is balanced by the accelerating force, limiting the acceleration to the saturated Lorentz factor
\begin{eqnarray}
  \gamma_{sat}&=&\left(\frac{3E_{||}\rho_c^2}{2e}\right)^{1/4} 
  \sim  10^7\left(\frac{f_r}{0.3}\right)^{1/4}\left(\frac{\mu_{WD}}{10^{35}~{\rm G\,cm^3}}\right)^{1/4} \nonumber \\
  &\times&\left(\frac{P}{500~{\rm s}}\right)^{-1/2}\left(\frac{\rho_c}{10^{9}~{\rm cm}}
  \right)^{1/4},
  \label{gsat}
\end{eqnarray}
where we adopt an electric field of $|E_{||}|\sim f_r\Phi_{a}/\rho_c$. Consequently, electrons can reach $\gamma\gtrsim10^7$ provided that $P<500$~s and $\mu_{WD}>10^{35}~{\rm G\,cm^3}$. The corresponding characteristic energy of curvature photons is $E_{c}\sim 100m_ec^2 (\gamma/10^7)^3(\rho_c/10^9~{\rm cm})^{-1}$.

  Within the framework of the present one-dimensional treatment, we 
  evaluate the gap size under the assumption of  the saturated motion,
  as discussed in Appendix~A. Alternatively, the gap size under the assumption of the unsaturated motion can be calculated by following the treatment of  \citet{2015ApJ...810..144T}. Comparing the gap size for the two regimes,
  the physically realized solution will correspond to the smaller gap size.
  By equating the gap sizes for two cases, we find that the saturation motion
  becomes important for the magnetic field less than $B\sim 8\alpha_f\chi_m B_c/3\sim 5\times 10^{10}(\chi_m/0.05)$~G with $\alpha_f\sim 1/137$ is the fine-structure constant and $B_c\equiv m_e^2c^3/(\hbar e) \sim 4.4\times 10^{13}$~G the critical quantum magnetic field.  The parameter $\chi_m$ is defined in section~\ref{sec:photon_absorption}.
  This boundary magnetic field is independent of the curvature radius and the spin period of the star.
  Since the typical surface magnetic field of WDs is below this boundary, the saturated particle motion should be taken into account.

\subsection{Magnetic pair creation process }
\label{sec:photon_absorption}
In this study, we analyze magnetic pair creation of gamma rays in strong magnetic fields. The opacity for magnetic pair creation may be expressed as~\citep{1966RvMP...38..626E,2006RPPh...69.2631H}
\begin{eqnarray}
    \alpha_{B}(E_\gamma,\theta_\gamma) &=& 0.23 \frac{\alpha_f}{\bar{\lambda}_\mathrm{c}} \frac{B}{B_c} \sin \theta_{\gamma} \exp{\left(-\frac{4}{3\chi}\right)} \nonumber \\ 
 &\sim& 990\left(\frac{B}{10^{9}~{\rm G}}\right)\sin \theta_{\gamma} \exp{\left(-\frac{4}{3\chi}\right)}~{\rm cm^{-1}}
\label{opacity}
\end{eqnarray}
where $\bar{\lambda}_\mathrm{c}=\hbar/m_ec$ is the reduced Compton wavelength of the electron, and $\theta_\gamma$ is the angle between the photon momentum and the local magnetic field. In our model, photons produced by curvature radiation undergo magnetic pair creation and are emitted tangentially to the magnetic field lines. As a photon propagates, the angle $\theta_{\gamma}$ gradually increases. To approximately describe the evolution of this angle as a function of propagation distance, we adopt the relation $\sin\theta_\gamma\sim \ell/(\rho_c^2+\ell^2)^{1/2}$, where $\ell=r-r_{em}$ and $r_{em}$ is the radial distance of the emission point. This expression is appropriate when the magnetic field lines are approximated as concentric circles with curvature radius $\rho_c$. In addition, $\chi$ appearing in equation~(\ref{opacity}) is defined as
\begin{equation}
    \chi \equiv  \frac{E_\gamma}{2m_ec^2} \frac{B}{B_c} \sin\theta_\gamma.
    \label{chi}
\end{equation}

\begin{figure*}
    
    \centering
    \subfigure[$\chi_m$ for  $E_\gamma = 10^3 m_e c^2$]{
    \includegraphics[scale=0.3]{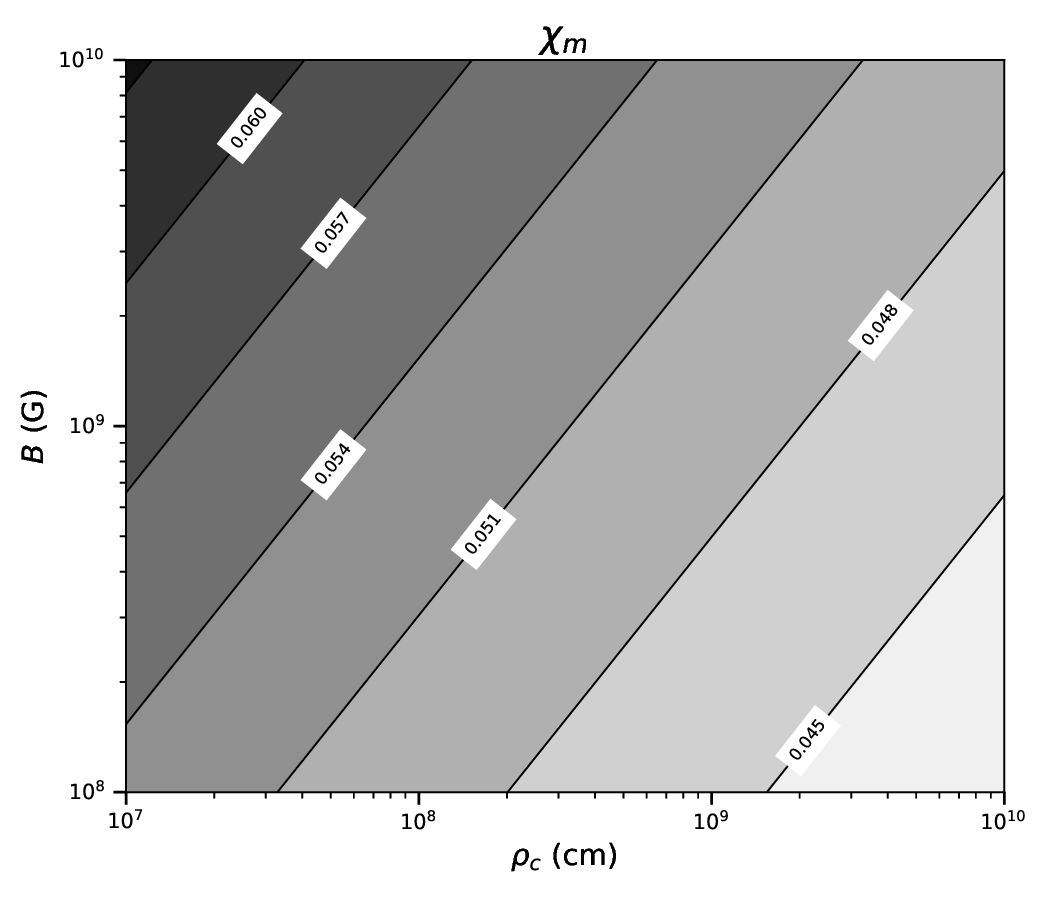}
    }
    \quad
    \subfigure[$\chi_m$ for  $E_\gamma = 10^4 m_e c^2$]{
    \includegraphics[scale=0.3]{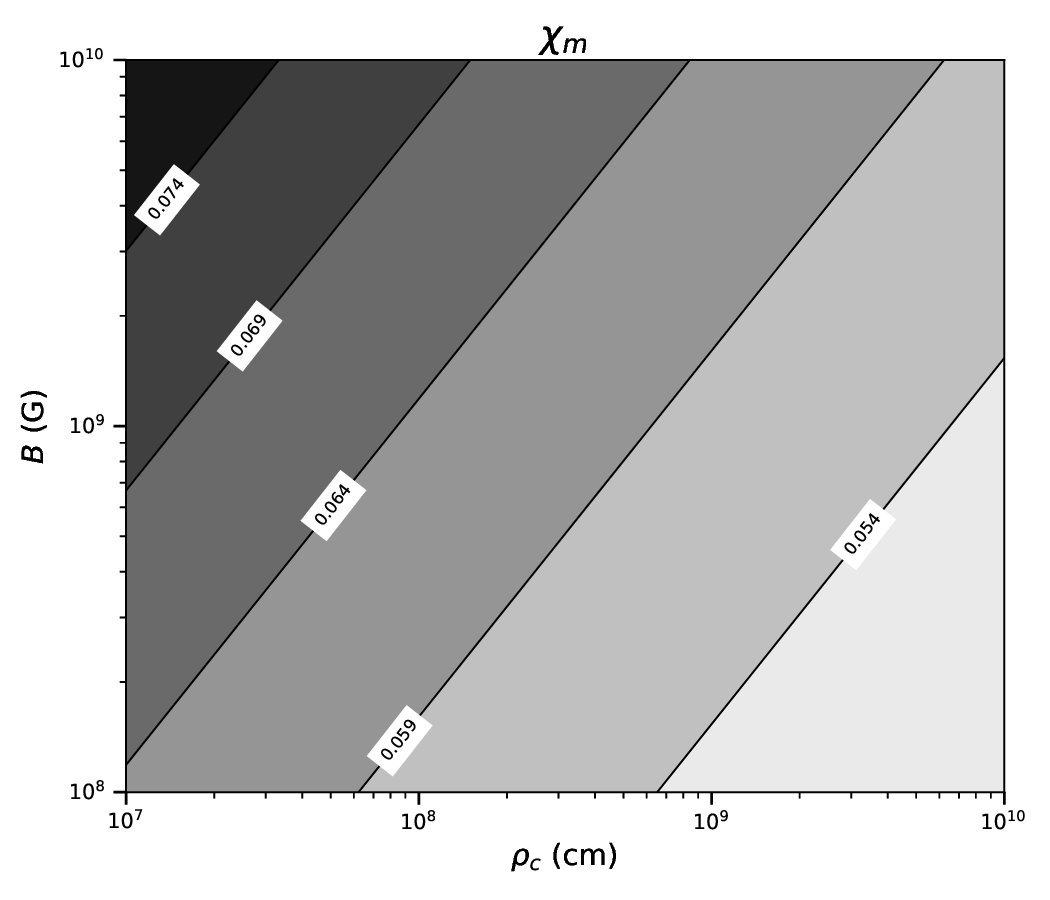}
    }
    \\
    \subfigure[$\ell_m$ for  $E_\gamma = 10^3 m_e c^2$]{
        \includegraphics[scale=0.3]{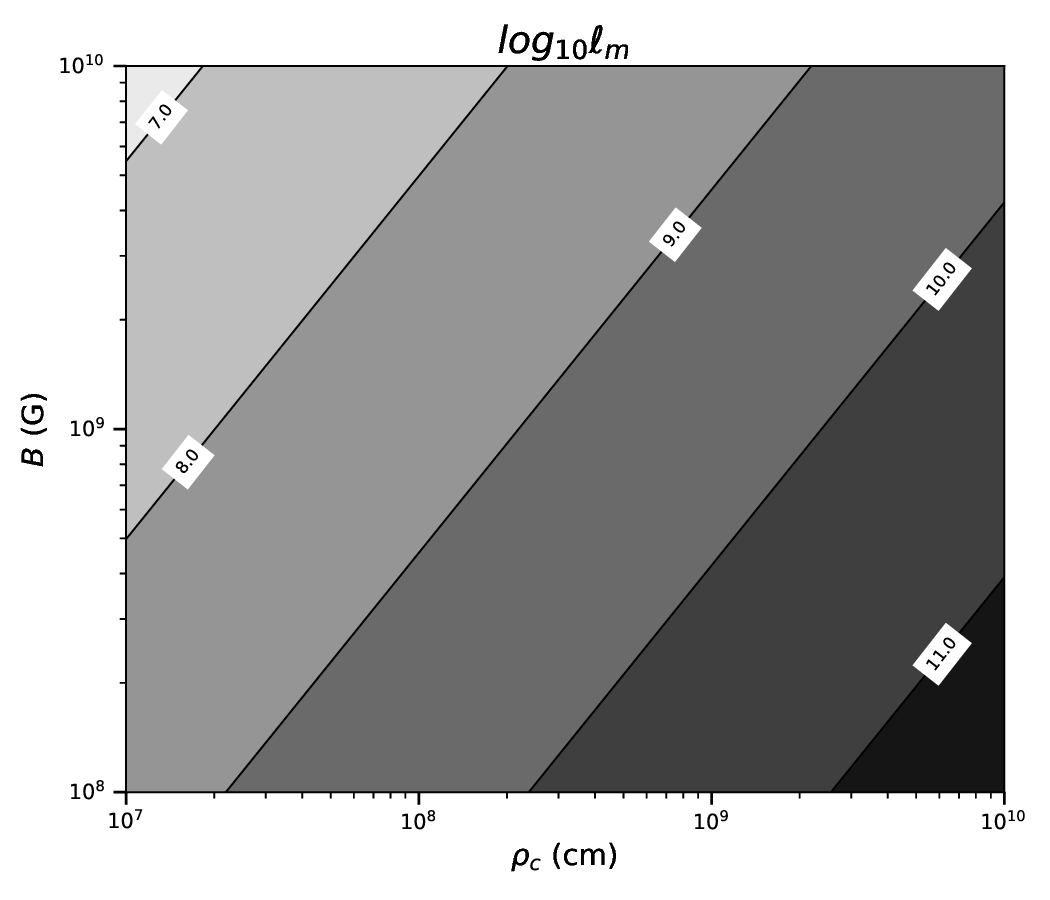}
    }
    \quad
    \subfigure[$\ell_m$ for $E_\gamma = 10^4 m_e c^2$]{
        \includegraphics[scale=0.3]{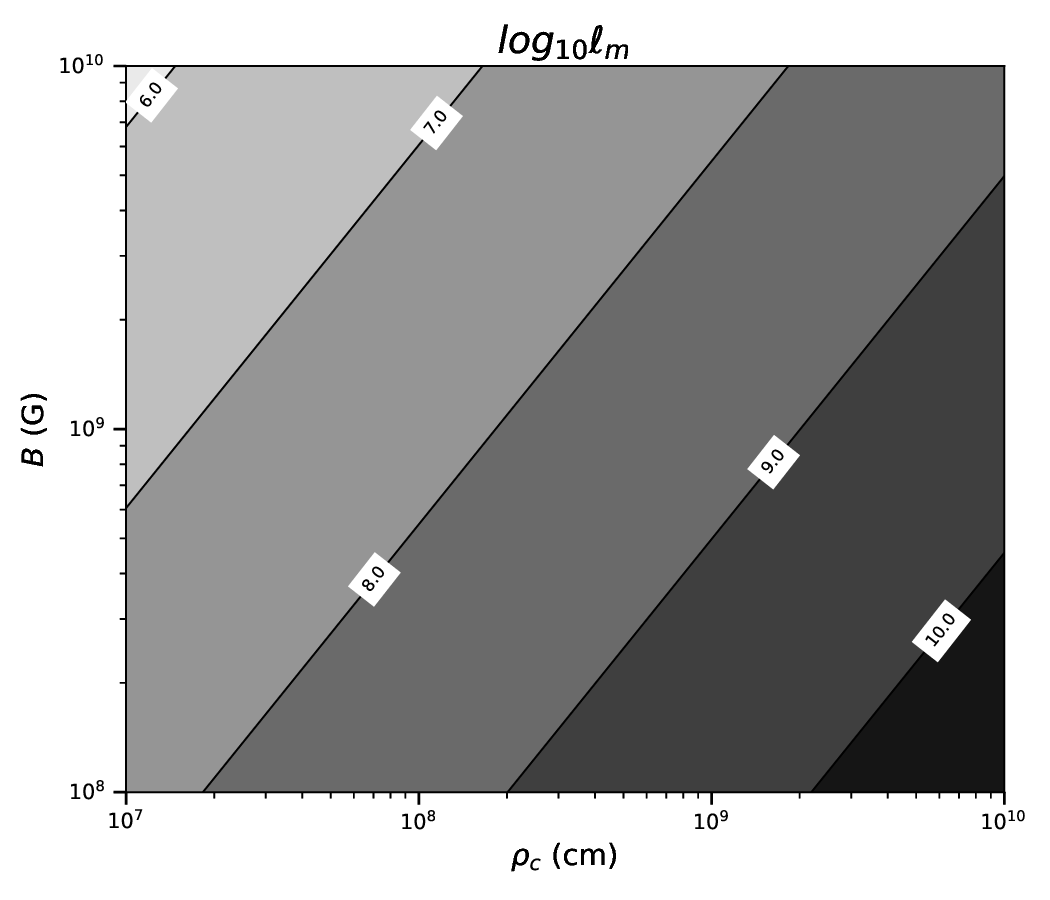}
    }
    \caption{Top panels: $\chi_m$. Bottom panels: the corresponding MFP $\ell_m$. The left and right panels show the results for photon energies of $E_{\gamma}=10^3\,m_ec^2$ and   
    $10^4\,m_ec^2$, respectively.}
    \label{fig_chi}
\end{figure*}

The optical depth $\tau$ is obtained by integrating the opacity $\alpha_B$ along the photon path. In the limit $\rho_c\gg \ell$ (so that $\sin\theta_{\gamma}\simeq \ell/\rho_c$), the optical depth becomes
\begin{eqnarray}
\tau&=&\int_0^\ell \alpha_B {\rm d}\ell\sim 8.7\times 10^4\left(\frac{\rho_c}{10^{9}~{\rm cm}}\right)\left(\frac{E_{\gamma}}{10^4m_ec^2}\right)^{-2}  \nonumber \\
&\times& \left(\frac{B}{10^9~{\rm G}}\right)^{-1} \int_{0}^{\chi(E_{\gamma})}
    \chi\exp\left(-{\frac{4}{3\chi}}\right)\mathrm{d}\chi,
  \label{eq:opt} 
\end{eqnarray}
where we assume a constant magnetic-field strength. In our numerical calculations, however, we take into account the variation of the magnetic-field strength along the photon path. We define $\chi_m(E_\gamma)$ by requiring that the optical depth reaches unity, i.e., $\tau(\chi_m)=1$. The corresponding MFP is then $\ell_m=\rho_c\chi_m(2m_ec^2/E_{\gamma})(B_c/B)$ from equation~(\ref{chi}). Figure~\ref{fig_chi} summarizes $\chi_m$ (top panels) and the resultant MFP $\ell_m$ (bottom panels) for photon energies of $E_{\gamma}=10^3m_ec^2$ (left panels) and $10^4m_ec^2$ (right panels). From the figure, we find $\chi_m\sim 0.05$ for the parameters $\rho_c\sim 10^{7-9}$~cm and $B\sim 10^{8-9}$~G relevant to magnetic WDs. Consequently, the MFP can be scaled as
\begin{eqnarray}
  \ell_m (E_\gamma)&=&2\rho_c\chi_{m}\frac{B_c}{B}\frac{m_ec^2}{E_{\gamma}}
  \sim 4.4\times 10^{8}~{\rm cm}\left(\frac{\chi_m}{0.05}\right) \nonumber \\
&\times&\left(\frac{\rho_c}{10^9~{\rm cm}}\right)
  \left(\frac{B}{10^{9}~{\rm G}}\right)^{-1}\left(\frac{E_{\gamma}}{10^4m_ec^2}\right)^{-1}.
  \label{mfp}
  \end{eqnarray}
These results suggest that magnetic WDs with surface fields $\gtrsim 10^9$~G can undergo magnetic pair creation efficiently when photons with $E_\gamma>10^4m_ec^2$ are produced near the stellar surface.

\subsection{Synchrotron radiation from generated pairs}

If electron--positron pairs are generated through magnetic pair creation, they are produced with a momentum component perpendicular to the magnetic field. This perpendicular component is rapidly lost via synchrotron radiation. Applying (i) $\gamma_{\pm}=E_{\gamma}/(2m_ec^2)$ as the initial Lorentz factor and (ii) $\sin\alpha=(B_c/B)(2m_ec^2/E_{\gamma})\chi_m$ as the sine of the pitch angle, the typical energy of synchrotron photons emitted by the pairs becomes
\begin{equation}
  E_{\gamma,{\rm syn}}=\frac{3\hbar e\gamma_{\pm}^2B\sin\alpha}{2 m_ec}\approx 0.04\left(\frac{\chi_m}{0.05}\right)E_{\gamma}.
\label{eq:esyn}
\end{equation}

After losing the perpendicular momentum through synchrotron radiation, the remaining energy of the pairs may be expressed as~\citep{2015ApJ...810..144T}
\begin{equation}
    E_{\mathrm{F}}={\left(1\:-\:\frac{v_{\parallel}^{2}}{c^2}\right)}^{-\frac{1}{2}} {m_e c^2}
    \approx \gamma_{\pm}m_ec^2[1+(\gamma_{\pm}\sin\alpha)^2]^{-1/2},
\end{equation}
where $v_{||}$ is the velocity component parallel to the magnetic field line and $v/c\approx 1$ has been used. Hence, the energy released through synchrotron radiation may be expressed as
\begin{equation}
  E_{\rm lost}=\gamma_{\pm}m_ec^2\left\{1-[1+(\gamma_{\pm}\sin\alpha)^2]^{-1/2}\right\}.
\label{elost}
\end{equation}
We note that, since $\gamma_{\pm}\sin\alpha\sim \chi_mB_{c}/B \gg 1$ for WDs, most of the initial energy of the pairs is lost through synchrotron radiation, i.e., $E_{{\rm lost}}\sim \gamma_{\pm}m_ec^2$.

\subsection{Calculation method}
\label{method}
We model electron acceleration, radiation, and pair creation within $R_{\rm WD}<r<2R_{\rm WD}$ using a one-dimensional grid along a representative open field line. In this paper, we apply $R_{\rm WD}=5\times 10^8$~cm unless otherwise stated.  For a given magnetic configuration, we calculate $\Phi_{\rm nco}$ from equation~(\ref{phimax}) for the dipole field or from equation~(\ref{q+d phi}) for the dipole-plus-quadrupole field, and evolve the Lorentz factor of the primary particles by solving equation~(\ref{eq:gam}). At each grid point, we evaluate the characteristic curvature-photon energy (equation~(\ref{ec})) and the number of emitted photons,
\[
\Delta N_{\gamma}=\frac{P_c}{E_c}\frac{\Delta s}{c}.
\]
For each curvature photon emitted at radial distance $r_{\rm em}$, we numerically integrate the opacity given by equation~(\ref{opacity}) along the photon trajectory and define the distance at which the optical depth reaches unity as the magnetic pair creation MFP $\ell_{\rm m}$. If the conversion radius $r_{\rm em}+\ell_{\rm m}$ lies within $2R_{\rm WD}$, we place the pair at the grid point closest to the conversion radius. If $r_{\rm em}+\ell_{\rm m}\ge 2R_{\rm WD}$, we assume that a fraction $1-e^{-\tau}\simeq\tau$ (for $\tau\ll1$) of a pair is created per photon, where $\tau$ is the optical depth integrated from $r_{\rm em}$ to $2R_{\rm WD}$, and include this contribution in the pair creation multiplicity. Although the contribution from the second case is minor compared to that from the first, we include it in order to reduce the numerical error in the multiplicity near the death line.

Newly created pairs radiate synchrotron photons with a total energy of $2\Delta N_{\gamma}E_{\rm lost}$ (equation~(\ref{elost})). We divide the synchrotron spectrum into six logarithmic bins and evaluate the number of photons in each bin using the standard single-particle synchrotron spectrum. Because $E_{\gamma,{\rm syn}}\simeq 0.04(\chi_m/0.05)E_{\gamma}$ (equation~(\ref{eq:esyn})), most synchrotron photons may escape without undergoing further magnetic pair creation. In Appendix~\ref{death}, we estimate the boundary in the $P$--$B_{*,d}$ plane above which pair creation by synchrotron photons dominates the pair multiplicity.

We do not model the detailed pair induced screening of the accelerating electric field. Instead, we approximate the screening by introducing a characteristic screening radius. For each grid point in $R_{WD}<r<2R_{WD}$, we calculate the conversion radius $r_{\rm em}+\ell_{\rm m}$ while neglecting the screening process. We then define the screening radius as
\begin{equation}
r_{\rm screen}\equiv \min\left[r_{\rm em}+\ell_{\rm m}(r_{em})\right],
\end{equation}
and assume $E_{\parallel}=0$ for $r>r_{\rm screen}$. Within this framework, the death line is defined by $r_{\rm screen}=2R_{WD}$ (Appendix~\ref{death}). Following the above procedure, we compute the pair multiplicity $\kappa$, defined as the number of pairs produced per primary electron, and evaluate the death line for WDs.

\section{result}
\label{sec:result}

\subsection{Dipole field configuration}
\begin{figure*}[htbp]
    \centering
    \subfigure{
    \includegraphics[scale=0.45]{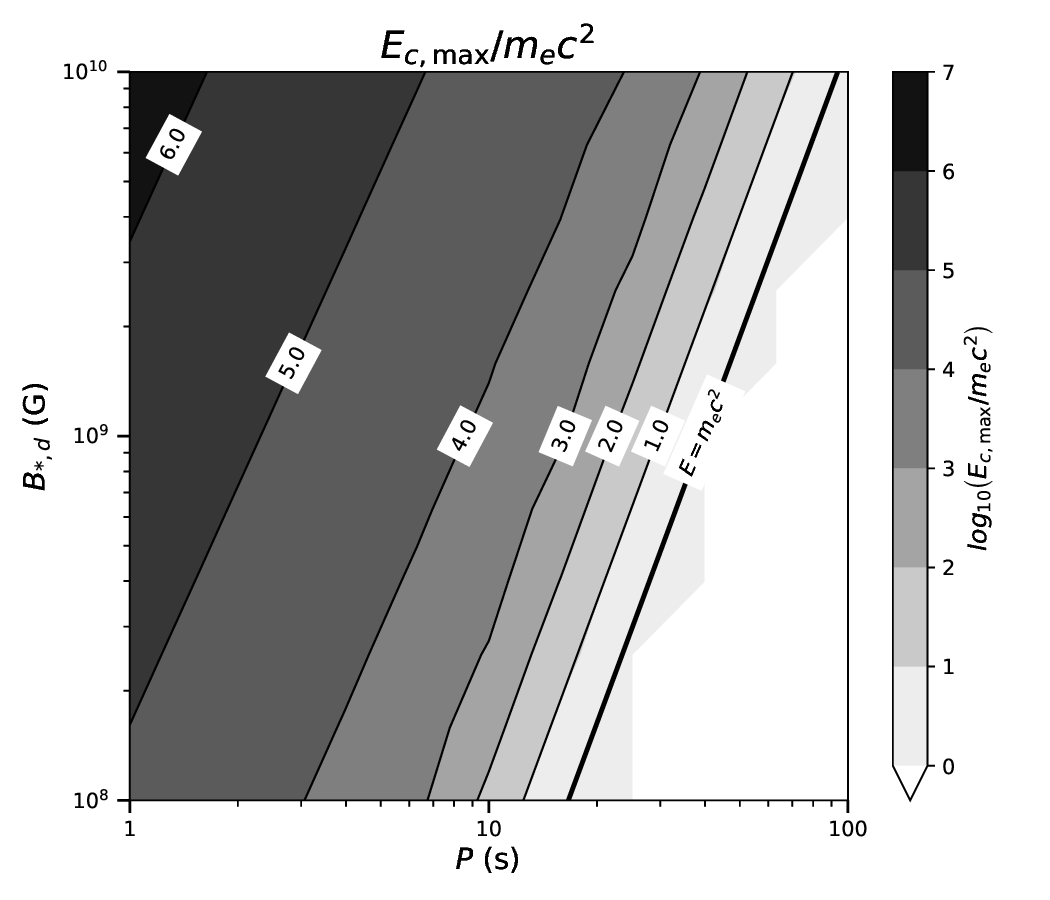}
    }
    \quad
    \subfigure{
    \includegraphics[scale=0.45]{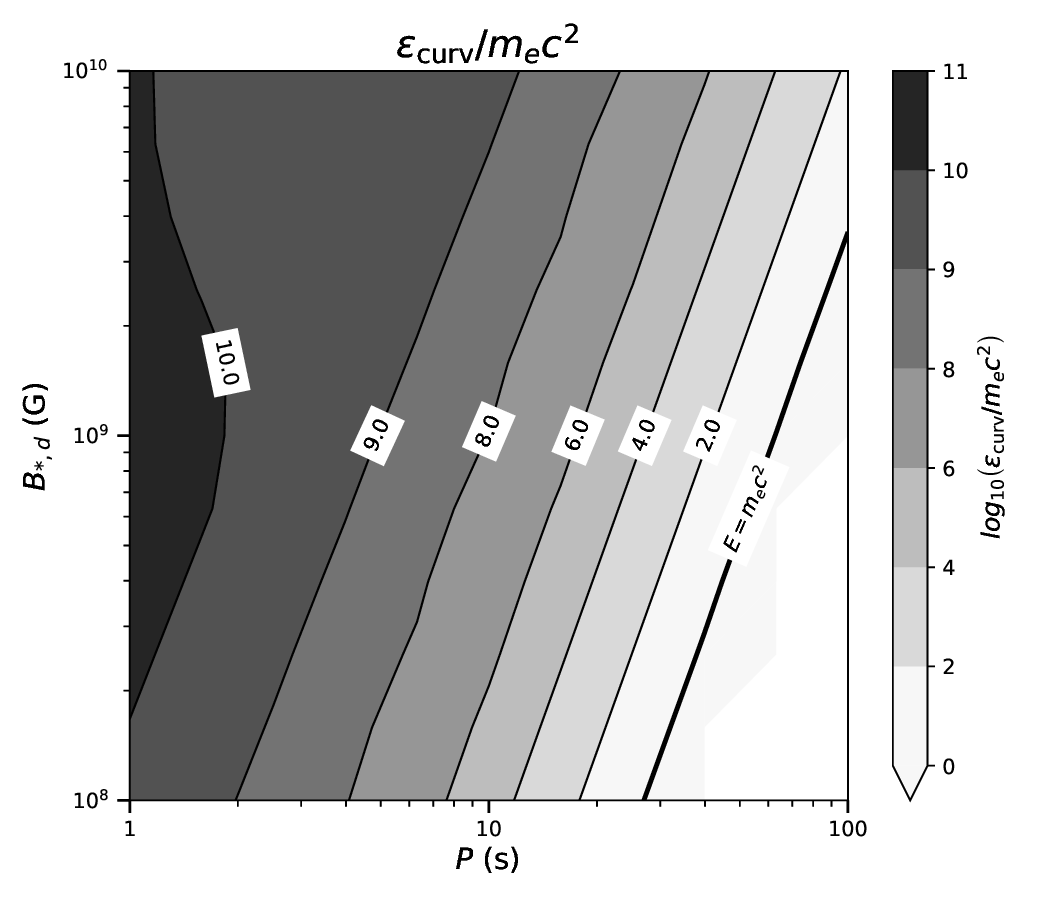}
    }
    \caption{Left: Maximum characteristic energy of curvature photons in $R_{WD}<r<2R_{WD}$. Right: Total energy of curvature radiation emitted by a single accelerated electron. The results are shown for a dipole-field geometry.}
    \label{curv}
\end{figure*}

Figure~\ref{curv} illustrates how the properties of curvature radiation above the WD polar cap depend on the spin period and the surface dipole magnetic field. The left panel shows the maximum characteristic energy of curvature photons produced within the calculated acceleration region, while the right panel shows the total radiated energy per primary electron,
\begin{equation}
\epsilon_{\rm curv}=\int_0^{\ell_{\rm gap}} P_c\,{\rm d}s/c,
\label{ecurv}
\end{equation}
where $\ell_{\rm gap}\sim r_{\rm screen}-R_{\rm WD}$ corresponds to the size of the acceleration region.
 The number of curvature photons emitted by a primary electron can be estimated as
  \begin{eqnarray}
  N_{curv}&\sim &\frac{P_{c}\ell_{gap}}{E_cc}
  \sim 2\times 10^3f_r^{-1/2}\left(\frac{\chi_m}{0.05}\right) \left(\frac{\rho_c}{R_{WD}}\right)^{1/2}
  \nonumber \\
  &&  \left(\frac{R_{WD}}{5\cdot 10^8{\rm cm}}\right)^{-1}\left(\frac{P}{1{\rm s}}\right)\left(\frac{B_{*,d\
}}{10^9{\rm G}}\right)^{-3/2},
\label{ncurv}
  \end{eqnarray}
where $\ell_{gap}\sim (m_ec^2/e|E_{||}|)(3\rho_c^2|E_{||}/2e)^{1/4}(8\chi_m\alpha B_c/9B_{*,d})$
is adopted~(Appendix~\ref{death}).  If the number of electrons extracted from the surface per unit time is of the order of the GJ value,
$\dot{N}_{\rm GJ}\sim 2\pi^2 B_{*,d} R_{\rm WD}^3/(P^2 e c)$,
the total curvature-radiation luminosity can be estimated as
\begin{eqnarray}
L_c &\sim& \dot{N}_{\rm GJ}\epsilon_{\rm curv} \sim
1.5\times10^{31}~{\rm erg~s^{-1}} \nonumber \\
&\times &
\left(\frac{P}{10~{\rm s}}\right)^{-2}
\left(\frac{B_{*,d}}{10^9~{\rm G}}\right)
\left(\frac{R_{\rm WD}}{5\cdot10^8~{\rm cm}}\right)^3
\left(\frac{\epsilon_{\rm curv}}{10^4 m_e c^2}\right).
\end{eqnarray}

As shown in Figure~\ref{curv}, the contour interval increases toward shorter spin periods. This is because, for sufficiently short periods, curvature-radiation drag limits particle acceleration to the radiation-saturated Lorentz factor (equation~(\ref{gsat})). As a result, both the maximum photon energy and the total radiated energy increase only slowly with decreasing $P$. In the right panel, the contour behavior changes around $\epsilon_{\rm curv}/m_ec^2\sim 10^{9}$. This transition reflects whether pair creation occurs efficiently within $r<2R_{WD}$, or equivalently whether the characteristic screening radius $r_{\rm screen}$ is smaller than $2R_{WD}$. For the region with $\epsilon_{\rm curv}/m_ec^2<10^{9}$ in the figure, the screening radius $r_{\rm screen}$ exceeds $2R_{WD}$, meaning that pairs are not produced within $r<2R_{WD}$. In this regime, the total radiated energy shown in the figure is estimated using $\ell_{\rm gap}\sim R_{WD}$ in equation~(\ref{ecurv}). In contrast, for the region with $\epsilon_{\rm curv}/m_ec^2>10^{9}$, the screening radius lies within $2R_{WD}$, or equivalently $\ell_{\rm gap}<R_{WD}$, and the energy available to the electron is limited by the onset of screening.

\begin{figure}
    \centering
    \includegraphics[width=1\linewidth]{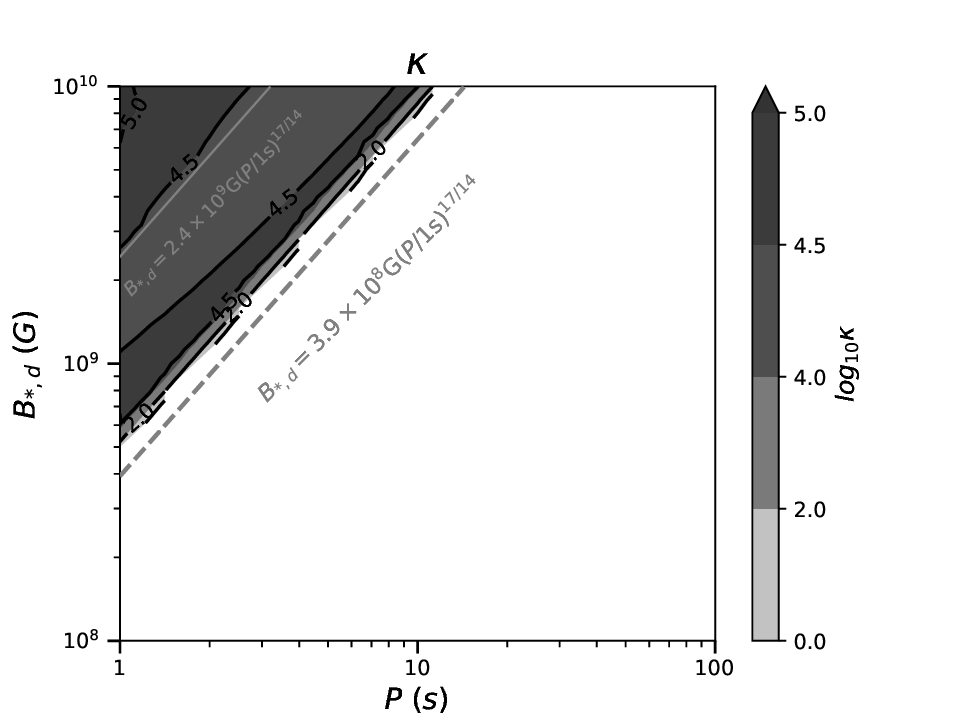}
    \caption{Contour plots of the pair multiplicity $\kappa$ for the dipole-field configuration. The dashed line represents the approximate expression (equation~(\ref{death1})) for the death line of the pair creation process, and the solid line shows the approximate boundary (equation~(\ref{B_syn1})) above which synchrotron photons dominate the pair cascade.}
    \label{multi}
\end{figure}

Figure~\ref{multi} illustrates how the pair creation multiplicity ($\kappa$) depends on $P$ and $B_{*,d}$. These results indicate that in a pure-dipole configuration, efficient pair production occurs only for very fast rotators ($P\sim1$~s), such as those expected from remnants of double-WD mergers~\citep{2011PhRvD..83b3002K,2025MNRAS.539.3013C}. In the current model, the approximate death line for magnetic pair creation in the dipole configuration is given by (Appendix~\ref{death})

\begin{equation}
  B_{*,d}\sim 3.9\times 10^8~{\rm G}
  \left( \frac{R_{\rm WD}}{5\times10^8~{\rm cm}}\right)^{-3/2}
  \left(\frac{P}{1~{\rm s}}\right)^{17/14},
\label{death1}
\end{equation}
(dashed line in Figure~\ref{multi}). The discrepancy between the numerical and analytical boundaries mainly arises from the treatment of the magnetic-field strength when evaluating the MFP. In the analytical calculation, we assume a spatially averaged magnetic-field strength within the acceleration region. In contrast, the numerical calculation accounts for the variation of the magnetic-field strength along the photon propagation path. The thin dark line of $B_{*,d}=2.4\times 10^9~{\rm G}(P/1~{\rm s})^{17/14}$ in the figure presents the boundary above which synchrotron photons emitted by the pairs are efficiently converted into pairs.

\subsection{Combination of dipole and quadrupole fields}
\begin{figure*}[htbp]
    \centering
    \subfigure{
        \includegraphics[width=0.45\textwidth]{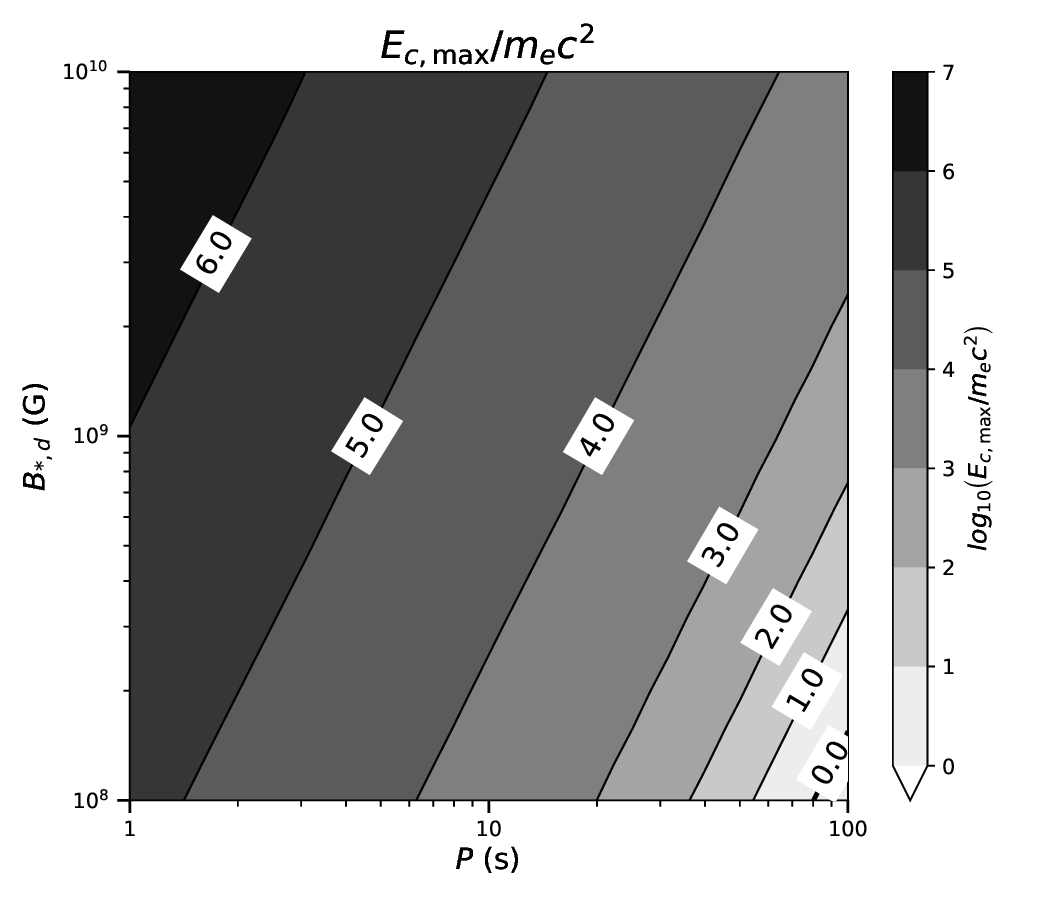}
    }
    \hfill
    \subfigure{
        \includegraphics[width=0.45\textwidth]{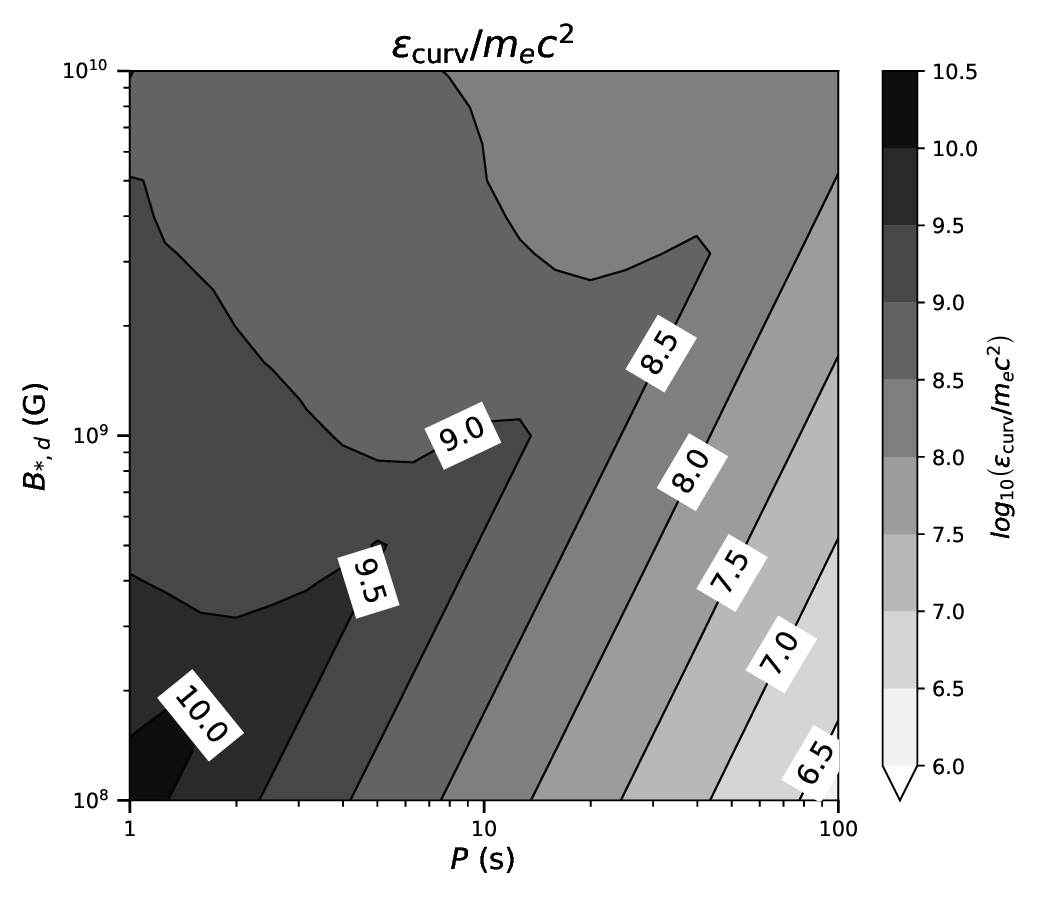}
    }
    \caption{Same as Figure~\ref{curv}, but for a combination of dipole and quadrupole fields with a field-strength ratio of $\beta=5$.}
    \label{fig:nonenergy}
\end{figure*}

This section presents results for a magnetic configuration consisting of a dipole field and a quadrupole component, although the actual surface-field structure may be more complex. We assume that (i) the dipole axis is aligned with the spin axis and (ii) the angle between the dipole and quadrupole axes is $\zeta\sim \pi/2$, which maximizes the potential drop near the stellar surface (Figure~\ref{pratio}). Figure~\ref{fig:nonenergy}, analogous to Figure~\ref{curv} for the pure-dipole case, summarizes the curvature-radiation properties for a field-strength ratio of $B_{*,q}/B_{*,d}=5$. Comparing the two magnetic configurations, we find that the influence of the quadrupole component becomes increasingly important at longer spin periods. In a pure-dipole geometry, the curvature radius near the polar cap scales as $\rho_c\sim (R_{\rm WD}R_{\rm lc})^{1/2}\simeq 5\times10^9(P/10~{\rm s})^{1/2}$~cm, which increases with $P$ and suppresses curvature radiation. In contrast, when the quadrupole component dominates over the dipole field, the curvature radius is reduced to $\rho_c\sim10^9$~cm (Figure~\ref{pratio}) and becomes much less sensitive to the spin period. Since the potential drop within $R_{\rm WD}<r<2R_{\rm WD}$ can be approximated as $\Phi_{nco}\sim(0.2$--$0.3)\Phi_a(R_{\rm WD}/\rho_c)$ and the curvature-photon energy scales as $E_c\propto\rho_c^{-1/4}$ for the saturated Lorentz factor, the presence of a strong quadrupole field leads to a substantial enhancement of curvature radiation at longer spin periods.

\begin{figure*}[htbp]
    \centering
    \includegraphics[width=0.47\textwidth]{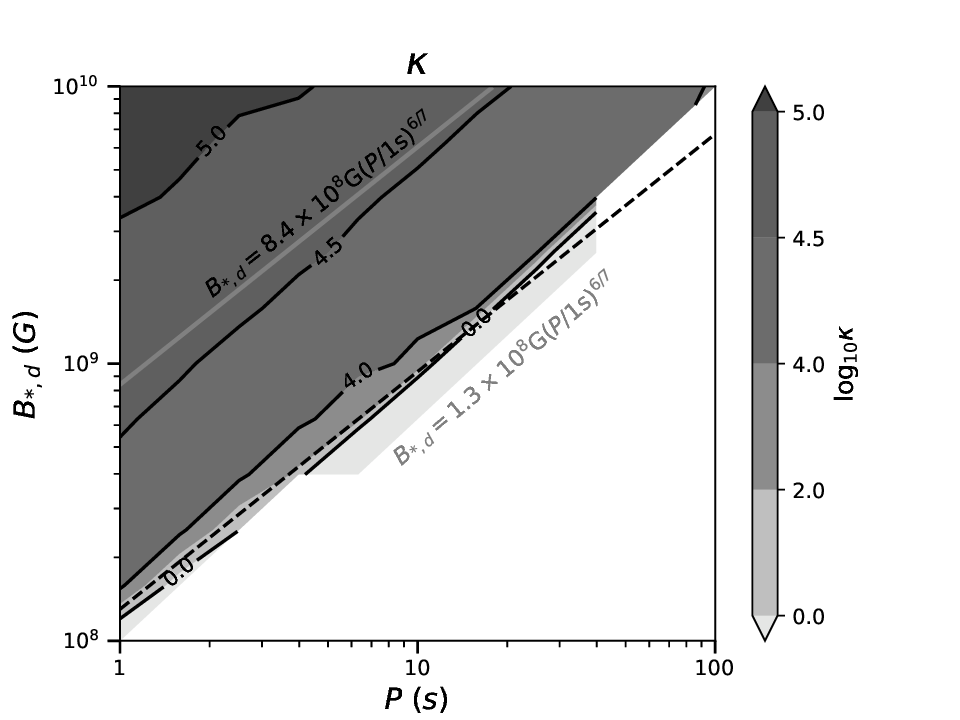}
    \includegraphics[width=0.45\textwidth]{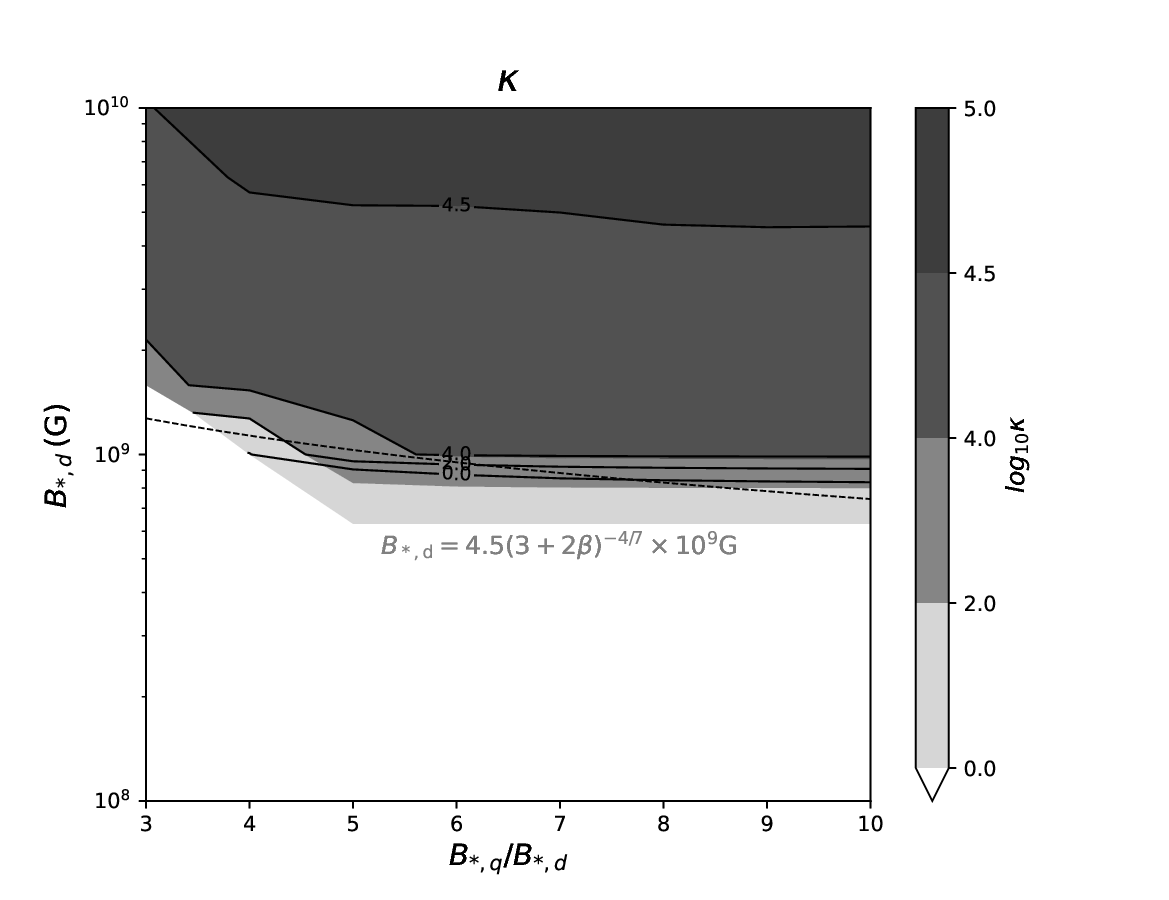}
    \caption{Pair multiplicity for the dipole--quadrupole configuration. Left: $\kappa$ on the $P$--$B_{*,d}$ plane for a fixed field-strength ratio of $\beta=5$. Right: $\kappa$ on the $\beta$--$B_{*,d}$ plane for a fixed period of $P=10$~s. The dashed lines show the death lines from equation~(\ref{death2}) with $f_r=0.2$ and $\rho_c=10^9$~cm. The solid line in the left panel shows the approximate boundary (equation~(\ref{B_syn2})) above w hich synchrotron photons dominate the pair cascade.}
    \label{multi_non}
\end{figure*}

Figure~\ref{multi_non} summarizes the pair creation multiplicity for the dipole--quadrupole configuration. The left panel shows $\kappa$ on the $P$--$B_{*,d}$ plane for a fixed field-strength ratio $\beta=5$, while the right panel displays $\kappa$ on the $\beta$--$B_{*,d}$ plane for a fixed spin period of $P=10$~s. Compared with Figure~\ref{multi}, this figure demonstrates that the presence of a quadrupole component significantly enhances magnetic pair creation. The enhancement becomes more pronounced at longer spin periods. For example, at $P\sim 2$~s, the minimum dipole-field strength required for pair creation is $B_{*,d}\sim 4\times 10^8$~G in the pure-dipole configuration, but decreases to $\sim 2\times 10^8$~G when both dipole and quadrupole fields are present. For a longer period of $P\sim 20$~s, the corresponding values are $B_{*,d}\sim 2\times 10^{10}$~G for the dipole case and $B_{*,d}\sim 10^9$~G for the combined configuration. This period dependence of the multiplicity enhancement is also related to the period dependence of the reduction in curvature radius, as discussed in the previous paragraph. For short periods, $\rho_c$ remains close to the dipole value, and the enhanced surface magnetic field due to the quadrupole component primarily drives the increase in multiplicity. For longer periods, however, the substantial reduction in $\rho_c$ leads to a larger potential drop and hence a stronger enhancement of pair creation. The death line (dashed lines in Figure~\ref{multi_non}) is evaluated as (Appendix~\ref{death})
\begin{eqnarray}
      B_{*,d}\sim & 6.2 (3+2\beta)^{-4/7} \times 10^8~{\rm G}
  \left(\frac{f_r}{0.2}\right)^{-3/7} \nonumber \\
  & \times \left(\frac{\rho_c}{10^9~{\rm cm}}\right)^{5/7}
    \left(\frac{P}{1~{\rm s}}\right)^{6/7}
  \left( \frac{R_{\rm WD}}{5\times10^8~{\rm cm}}\right)^{-13/7}.
\label{death2}
\end{eqnarray}
Figure~\ref{multi_non} indicates that the maximum multiplicity is of the order of $\kappa\sim 10^5$ for dipole fields with $B_{*,d}<10^{10}$~G.

\section{Discussion}
\label{sec:discussion}

\subsection{Comparison with previous studies for death line}
\citet{2011PhRvD..83b3002K} evaluated the death line of magnetic WDs as $B_{*,d}\sim 3\times 10^4~{\rm G}(P/1~{\rm s})^{13/8}(R_{WD}/5\cdot 10^{8}~{\rm cm})^{-19/8}$, assuming a curvature radius of $\rho_c\sim R_{WD}$. Compared with our estimate in equation~(\ref{death2}) for the dipole--quadrupole configuration, their death line allows pair creation at significantly longer spin periods for a given magnetic-field strength. This discrepancy primarily arises from the different assumption regarding the maximum Lorentz factor achievable by electrons in the polar-cap accelerator. While \citet{2011PhRvD..83b3002K} estimated the maximum Lorentz factor as $\gamma_{\max}=e\Phi_{a}/(m_ec^2)$, our results show that the saturated Lorentz factor $\gamma_{\rm sat}$ in equation~(\ref{gsat}) is more appropriate for evaluating the death line. Because $\gamma_{\rm sat}$ is substantially lower than $\gamma_{\max}$, our predicted death line imposes a more stringent constraint on the viability of magnetic pair creation in WD magnetospheres.

In contrast, \citet{2012JPhG...39f5001B} proposed a death line of $B_{*,d}=1.8\times 10^{10}~{\rm G}$, independent of the spin period, under the influence of energy losses due to curvature radiation. This threshold is considerably more restrictive than our prediction. This difference primarily arises from the estimate of the saturated Lorentz factor. \citet{2012JPhG...39f5001B} adopted the Lorentz factor obtained by balancing the cooling timescale of curvature radiation with the dynamical timescale, yielding $\gamma_{\rm sat}\sim 1.4\times 10^7$. This approach may be valid in the absence of an accelerating electric field. However, in the presence of an accelerating electric field, the saturated Lorentz factor can exceed $\gamma_{\rm sat}>10^7$ for $P<500$~s, as shown by equation~(\ref{gsat}). This suggests that our estimate, with a self-consistent treatment of the accelerating electric field, may provide a more realistic constraint on the death line of magnetic WDs.

 In the current model, the position of the death line for dipole-quadrupole
field configuration depends on the WD radius as $B_{*,d}\propto R_{WD}^{-8/7}$,
as equation~(\ref{death2}) with $\rho_c\propto R_{WD}$ indicates. This dependence is weaker than that obtained in previous estimates, primarily because the present model assumes a self-similar dipole--quadrupole
magnetic geometry with $\rho_c\propto R_{WD}$. We also note that
for the $B_{*,d}-P$ regions far from the death line (upper-left ward in Figure~\ref{multi_non}), most curvature photons emitted inside the accelerator are converted into pairs, which subsequently produce synchrotron photons. Hence, the multiplicity will be scaled to $\kappa\propto N_{curv}\propto R_{WD}^{-1}$ (equation~(\ref{ncurv})).

\subsection{Influence of inverse-Compton scattering}
\label{sec:IC}
Accelerated electrons may upscatter the stellar photons  via inverse Compton scattering (hereafter IC) and produce very-high-energy gamma rays.
Such gamma rays can subsequently be converted into pairs through magnetic
pair creation. Nevertheless, as we will show below,   the IC process provides only a minor impact on both  dynamics of the primary particles and the magnetic pair-creation process.

The ratio of the acceleration time ($t_a$) to the cooling timescale of
the IC in Thomson regime  ($t_{IC}$) is of the order of 
\begin{eqnarray}
  \frac{t_{a}}{t_{IC}}&\sim& \frac{(4\sigma_T c\gamma^2U_{ph}/3)}{eE_{||}c}
  \sim 6\times 10^{-3}\left(\frac{\gamma}{10^5}\right)^2\left(\frac{T_{WD}}{3\cdot 10^4~{\rm K}}\right)^4\nonumber \\
&&  \left(\frac{\rho_c}{R_{WD}}\right)\left(\frac{R_{WD}}{2.5\cdot 10^8~{\rm cm}}\right)^{-2} \left(\frac{P}{400~{\rm s}}\right)^2\left(\frac{B_{*,d}}{10^{9}~{\rm G}}\right)^{-1}
\label{timescale}
\end{eqnarray}
Here $\sigma_T$ is the Thomson cross section, $U_{\rm ph}\sim a_{\rm SB}T_{\rm WD}^4$ with $T_{WD}$ being the surface temperature is the characteristic blackbody energy density near the stellar surface  and $a_{\rm SB}=8\pi^5 k_{\rm B}^4/(15c^3h^3)$~\citep{RybickiLightman1979}.
For the rapidly rotating and massive WDs ZTF J2008+4449 and J1901+1458,
whose surface temperatures are approximately
$T_{\rm WD}\sim3\times10^4\,{\rm K}$, the IC process enters the Klein-Nishina regime at  $\gamma\sim 10^5$. In this regime, the cooling timescale increases with the Lorentz factor, further reducing the efficiency of the radiation losses. Therefore, equation~(\ref{timescale}) indicates that the IC cooling has a negligible effect  on the dynamics of the electrons, except for very hot WDs with $T_{WD}>10^5$~K.

The relative importance of IC emission compared with curvature radiation can be estimated as
\begin{eqnarray}
 \frac{P_{IC}}{P_c}&\sim& 10^{-3}\left(\frac{\gamma}{10^4}\right)^{-4}\left(\frac{T_{WD}}{3\cdot 10^4~{\rm K}}\right)^2\nonumber \\
&&\left(\frac{\rho_c}{R_{WD}}\right)^2\left(\frac{R_{WD}}{2.5\cdot 10^8~{\rm cm}}\right)^{2},
\label{pic}
\end{eqnarray}
where we adopted  $P_{IC}\sim \sigma_TcU_{ph}(m_ec^2/2.8k_BT_{WD})^2$ for the IC power in the Klein-Nishina regime. Equation~(\ref{pic}) shows
that the IC luminosity is substantially smaller than the curvature-radiation power for $T_{WD}\sim 3\times 10^4~{\rm K}$ of ZTF J2008+4449 and J1901+1458, suggesting the  IC photons are unlikely to contribute significantly to the pair-creation cascade in WD.
We therefore conclude that  unless $T_{\rm WD}\gtrsim10^5~{\rm K}$ of very hot young WDs, the IC process provides only a minor contribution to
the dynamics and pair-creation process of the polar cap accelerator.

\subsection{Influence of three-dimensional geometry}
\label{threeg}
 
As described in section~\ref{sec:model}, the present calculation adopts a one-dimensional treatment along the central field line of the magnetic flux tube of the polar cap model. For clarity  in this section, we  define the multiplicity that is  related to the pairs created
within the  {\it open} field line region, which will subsequently contribute to  observable multi-wavelength emission. The current one-dimensional treatment will overestimate this multiplicity compared with a  three-dimensional treatment for the following two reasons. 

First, when the mean-free path of the magnetic pair-creation is comparable or longer than the polar cap radius, the photon may escape the acceleration region through  the side boundary  of the flux tube before undergoing pair-creation process. To assess this effect, we estimate the
transverse displacement of an emitted photon using a simple geometrical relation, $\ell_{\perp}\sim\sqrt{\ell_{MFP}^2+\rho_c^2}-\rho_c$
and  evaluated the fraction of the photons emitted along the central  field line whose transverse displacement exceeds
the polar cap size, taken as the characteristic  transverse size of the flux tube. We found that this effect reduces
the pair multiplicity to approximately 65\% of the values obtained in our one-dimensional calculations. 

Second, the present model  assumes the maximum electric potential drop at the center of the flux tube~(equation~(\ref{q+d phi})). In reality, the magnitude of the potential  drop decreases toward the side of the polar cap and vanishes  at the last open field line. Consequently,  the
multiplicity averaged over the polar cap region is expected to be smaller than that obtained from the one-dimensional model. Since the potential drop is the quadratic function in the transverse-direction, we calculated the multiplicity by reducing the accelerating electric field by a
factor of two.  We found that the resulting  multiplicity is reduced by less than a factor of two, except in the vicinity of the
death-line. Combining these effects, we may conclude that for $B_{*,d}-P$ parameter space sufficiently  far from the death-line,
the influence of the three-dimensional geometry is unlikely  to alter the predicted multiplicity by more than an order of magnitude.

\subsection{X-ray emission from the isolated magnetic WDs}
Recent observations have detected X-ray emission from two isolated magnetic WDs, ZTF J2008+4449 and J1901+1458, whose X-ray luminosities in the 0.1--1~keV band are $L_X\sim 10^{29}~{\rm erg~s^{-1}}$ and $\sim 10^{27}~{\rm erg~s^{-1}}$, respectively~\citep{2026A&A...706A.188C,2021Natur.595...39C,2024PASJ...76..702B,2025arXiv250903216D}. These WDs have similar spin periods ($P\sim 390$~s for ZTF J2008+4449 and $\sim 416$~s for ZTF J1901+1458), surface magnetic fields of the order of $B\sim (0.5$--$1)\times 10^9$~G, and spin-down powers of $L_{\rm sd}\sim 5\times 10^{30}~{\rm erg~s^{-1}}$. The origin of their X-ray emission remains uncertain, with proposed mechanisms including accretion from a debris disk, magnetic interaction with the interstellar medium, and curvature radiation near the stellar surface.

For a pure-dipole field configuration, electrons are accelerated only to $\gamma\sim 2\times 10^4$ near the stellar surface, as estimated from equation~(\ref{gamma}) with the parameters of ZTF J1901+1458 ($P\sim 400$~s, $B_{*,d}\sim 10^9$~G and $R_{WD}\sim 2.5\times 10^8$~cm). The corresponding curvature radius of the field lines near the stellar surface is roughly $\rho_c\sim (R_{\rm WD}R_{\rm lc})^{1/2}\sim 2\times 10^{10}$~cm. Within the current SCLF framework, this yields a characteristic curvature-photon energy of only $E_c\sim 3\times 10^{-5}$~eV (equation~(\ref{ec})), which is far too low to explain the observed X-ray emission. As discussed in Section~\ref{nondipole}, introducing a quadrupole component reduces the curvature radius to $\rho_c\sim R_{WD}$ and enhances the potential drop, allowing the Lorentz factor to reach $\gamma\sim 5\times 10^6$ (equation~(\ref{gsat})). To produce X-rays via curvature radiation, the required Lorentz factor is $\gamma\sim 2\times 10^5(E_c/1~{\rm keV})^{1/3}$.  Although this value is achievable in a dipole--quadrupole configuration, the resulting luminosity is still far too low to explain the observations,
\begin{eqnarray}
  L_X&\sim &P_c\dot{N}_{\rm GJ}\frac{R_{\rm WD}}{c}\sim 10^{23}~{\rm erg~s^{-1}}\left(\frac{E_c}{1~{\rm keV}}\right) \nonumber \\
    &\times&\left(\frac{R_{WD}}{2.5\cdot 10^8~{\rm cm}}\right)^{10/3}\left(\frac{P}{400~{\rm s}}\right)^{-2}.
\end{eqnarray}
This discrepancy suggests that curvature radiation alone is unlikely to account for
  the observed X-ray luminosities of ZTF J2008+4449 and J1901+1458. It has been suggested that
  low-level accretion from a  debris disk, which might form through the fall back process after
  the double white dwarf merger or tidal disruption of a planetary body,  could provide a WD's  heating mechanism capable of
  producing $L_X\sim 10^{27-29}~{\rm erg~s^{-1}}$~\citep{2026A&A...706A.188C}. 
  Accretion from a debris disk surrounding a merger-formed magnetic WD has also been proposed as the origin of anomalous X-ray pulsars~\citep{2020ApJ...895...26B}.
  An alternative scenario, in which the WD outflow interacts with the debris disk, has also been proposed to explain the observed X-ray emission~\citep{2025arXiv250903216D}.

If electrons are accelerated to a maximum Lorentz factor of $\gamma\sim 10^7$ in the magnetospheres of ZTF J2008+4449 and J1901+1458, the resulting curvature radiation would produce emission in the $\sim 20$~MeV range with a luminosity of $L_{c}\sim 10^{30}~{\rm erg~s^{-1}}$. This emission would correspond to a flux of $F\sim 10^{-12}(d/100~{\rm pc})^{-2}~{\rm erg~cm^{-2}~s^{-1}}$, potentially detectable by future MeV missions.

\subsection{Implications for LPTs}
As illustrated in Figure~\ref{multi_non} and described by our death line in equation~(\ref{death2}), magnetic pair creation in WDs with typical dipole magnetic fields of $B_{*,d}=10^{8-9}$~G becomes viable only when the spin period is shorter than $P\sim10$~s. Within the framework of our model, therefore, if the observed radio periods of LPTs, which are typically of the order of hours, correspond to the spin periods of WDs, the expected accelerating potential falls significantly below the threshold required to sustain pair creation, unless the WD is extremely strongly magnetized. Even in the case of CHIME J0630+25, which has the shortest radio period ($P \sim 400$~s) among the LPTs, the required dipole magnetic field would be $B_{*,d}>10^{10}$~G. Our results suggest that, if LPTs are indeed powered by WDs, an alternative physical process, rather than the rotation-powered activity of {\it isolated} WDs, must be invoked. One possibility is magnetic interaction with a companion star located at $r\sim 10^{11}$~cm from the WD if the orbital period is several hours. Such an interaction could enhance the available potential drop at the WD surface beyond the standard expression in equation~(\ref{phimax}), thereby relaxing the spin-period threshold required to trigger pair creation.

An important theoretical uncertainty in the death-line estimate is how non-dipolar magnetic fields influence the curvature radius and the accelerating potential near the stellar surface. As shown in Figure~\ref{pratio}, a quadrupole component enhances the potential drop up to $\Phi_{nco}\sim 0.25\Phi_a$. If higher-order multipole fields could further increase the potential drop to $\Phi_{nco}\sim \Phi_a$, the dipole-field threshold for magnetic pair creation would be reduced to
$B_{*,d}\sim 2(9+6\beta)^{-4/7}\times 10^8(R_{WD}/5\cdot 10^8~{\rm cm})^{5/7}(P/1~{\rm s})^{6/7}$~G,
assuming $f_r=1$ and $\rho_c=R_{WD}$ in equation~(\ref{death2}).
This would make pair creation more feasible; for example,
$B_{*,d}\sim 4\times 10^9~{\rm G}$ with $\beta=5$ and $R_{\rm WD}=5\times 10^8$~cm
for CHIME J0630+25.

Such complex magnetic structures may naturally arise in massive,
highly magnetized WDs formed through double-WD mergers
~\citep{2026JHEAp..5300593M}. Recent studies suggest that merger remnants of double-WD mergers may produce detectable thermal optical transients, with potentially high detection rates in future wide-field surveys such as the Legacy Survey of Space and Time (LSST) conducted by the Vera C. Rubin Observatory~\citep{2023ApJ...958..134S}. If these newly formed WDs are born rapidly rotating and strongly magnetized, their rotation-powered activity may generate detectable non-thermal emission or additional heating signatures in the optical/IR light curves. Therefore, future multi-wavelength observations of double-WD merger remnants will provide an important opportunity to test the rotation-powered WD scenario discussed in this paper.

\section{Summary}
\label{summary}
We have discussed the rotation-powered activity of magnetic WDs and evaluated curvature radiation and magnetic pair creation within the framework of the SCLF model above the polar-cap region. We considered these processes for two types of magnetic-field configuration, namely, a pure dipole field and a combination of dipole and quadrupole fields. For a pure-dipole configuration, significant pair production is achieved only for extremely short spin periods and strong surface magnetic fields. In contrast, the inclusion of a quadrupole component can substantially reduce the curvature radius and enhance the accelerating potential near the stellar surface. As a result, the death line is shifted significantly toward longer spin periods. These results demonstrate that magnetic-field geometry plays a crucial role in determining whether magnetic pair creation can occur in WD magnetospheres. Compared with previous studies, we evaluate the death line for WDs with a more self-consistent treatment of the accelerating electric field.

\begin{acknowledgments}
  We thank the anonymous referee for detailed and  constructive suggestions for the manuscript.
  We appreciate Drs S. Kisaka and K. Kashiyama for useful discussion.
   W.B. and J.T. are supported by the National Key Research and Development Program of China (grant
No. 2020YFC2201400) and the National Natural Science Foundation of China
(grant No. 12573044).
\end{acknowledgments}

\appendix
\section{Death Line and Synchrotron-Dominated Boundary of Magnetic Pair Creation} 
\label{death}
\begin{figure}
    \centering
    \includegraphics[width=0.5\linewidth]{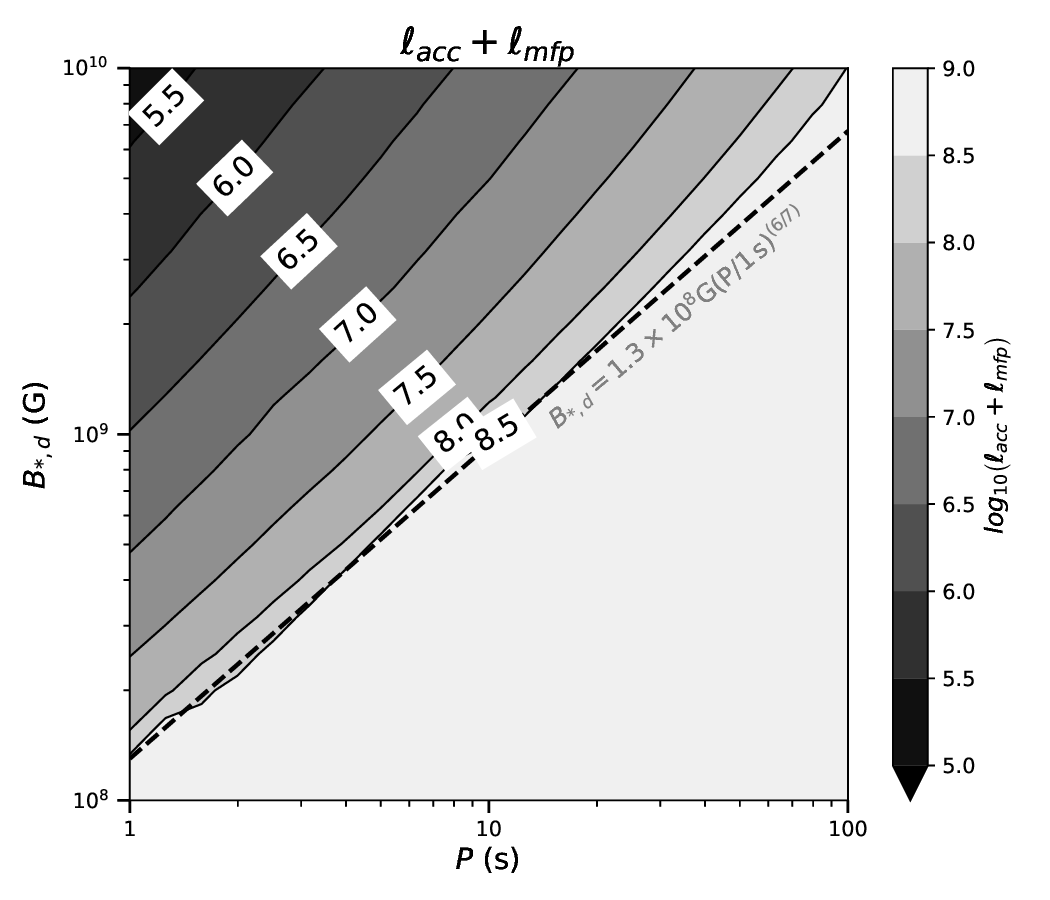}
    \caption{Contour plots of $\ell_{\rm acc}+\ell_{\rm m}$ on the $P$--$B_{*,d}$ plane. The results are shown for the dipole--quadrupole configuration with a field-strength ratio of $\beta=5$. The dashed line indicates the analytical death line for $f_r=0.2$.}
    \label{fig:nondipole_gap}
\end{figure}
In this section, we examine the death line for magnetic pair creation in WD magnetospheres within the framework of the SCLF model by applying the pair creation condition $r_{\rm em}+\ell_{\rm m}<2R_{\rm WD}$ discussed in Section~\ref{method}. In Section~\ref{method}, we evaluated the screening radius as
\begin{equation}
  r_{\rm screen}=\min\left[r_{\rm em}+\ell_{\rm m}(r_{\rm em})\right],
  \label{rs-append}
\end{equation}
where $r_{\rm em}$ is the emission radius and $\ell_{\rm m}$ is the MFP in the region $R_{\rm WD}<r_{\rm em}<2R_{\rm WD}$. In our model, the death line corresponds to $r_{\rm screen}=2R_{\rm WD}$. For WDs, we find that electrons whose motion is limited by the balance between the accelerating force and radiation-reaction drag emit curvature photons that are converted into pairs, as shown in Figures~\ref{curv} and~\ref{fig:nonenergy}. Hence, the screening radius is determined by the location where the electrons reach the saturated state. The radial distance scale required to reach saturation is $r_{\rm em}\sim R_{\rm WD}+\ell_{\rm acc}$, where $\ell_{\rm acc}\equiv \gamma_{\rm sat}m_ec^2/(e|E_{\parallel}|)$ with $\gamma_{\rm sat}=(3E_{||}\rho_c^2/2e)^{1/4}$ is the acceleration length measured from the stellar surface.

For a dipole-field configuration, the parallel electric field can be approximated as $|E_{||}|\sim 9\Phi_a/(32\rho_c)$, where $\rho_c\sim (R_{\rm WD}R_{\rm lc})^{1/2}$, as described in equation~(\ref{epara1}). For the non-dipolar field geometry, the electric field can reach $|E_{||}|\sim 0.25\Phi_a/\rho_c$ with $\rho_c\sim 10^9~{\rm cm}$, as illustrated in Figure~\ref{pratio}. Consequently, the pair creation condition~(\ref{rs-append}), assuming a constant magnetic field strength and curvature radius, can be written as
\begin{equation}
  \left(\frac{m_ec^2}{e|E_{||}|}\right)
  \left(\frac{3\rho_c^2|E_{||}|}{2e}\right)^{1/4}
  \left(1+\frac{8}{9}\chi_m\alpha \frac{B_c}{B}\right)<R_{\rm WD},
  \label{death-cond}
  \end{equation}
where $\alpha=e^2/\hbar c$ is the fine-structure constant. For the WD case with $\chi_m\gtrsim 0.05$ (Figure~\ref{fig_chi}), the first term in the parentheses can be neglected for $B<10^{10}$~G; equivalently, the screening scale is characterized by the MFP ($\ell_{\rm acc}<\ell_{\rm m}$).

To derive an approximate death line, we adopt a radially averaged description over the region $R_{\rm WD}<r<2R_{\rm WD}$. Since the magnetic pair creation opacity (equation~\ref{opacity}) depends on $B\sin\theta_\gamma$, the relevant quantity to average is $B\sin\theta_\gamma$ rather than the magnetic-field strength alone. As the photon propagates, the propagation angle increases approximately as $\sin\theta_\gamma \sim \ell/\rho_c$ for $\theta_\gamma \ll 1$. Accordingly, we approximate this quantity by its radial average,
\begin{eqnarray}
\langle B\sin\theta_\gamma \rangle 
&=& \frac{1}{R_{\rm WD}} \int_{R_{\rm WD}}^{2R_{\rm WD}} 
\left[B_d(r)+B_q(r)\right] \sin\theta_\gamma(r)\, dr \nonumber\\
&\simeq& B_{*,d}\left(\frac{1}{8}+\frac{1}{12}\beta\right)\sin\theta_\gamma(2R_{\rm WD}),
\end{eqnarray}
where we use $B_d(r)=B_{*,d}(R_{\rm WD}/r)^3$ and $B_q(r)=B_{*,q}(R_{\rm WD}/r)^4$. This approximation provides an order-of-magnitude estimate of the pair creation condition. For convenience, we define an effective magnetic-field strength
\begin{equation}
B_{\rm th}\equiv B_{*,d}\left(\frac{1}{8}+\frac{1}{12}\beta\right),
\end{equation}
so that the averaged quantity can be written as $\langle B\sin\theta_\gamma \rangle \simeq B_{\rm th}\sin\theta_\gamma(2R_{\rm WD})$. Substituting this expression into equation~\ref{death-cond}, we obtain the approximate death line. For a dipole-field configuration ($\beta=0$), the resulting death line is expressed as 
\begin{equation}
  B_{*,d}\sim 3.9\times 10^8~{\rm G}
  \left(\frac{R_{\rm WD}}{5\cdot10^8~{\rm cm}}\right)^{-3/2}
  \left(\frac{P}{1~{\rm s}}\right)^{17/14},
\label{appen-death1}
\end{equation}
where $\chi_m=0.05$ is adopted.

For the dipole--quadrupole configuration, we adopt $\rho_c\sim 10^9~{\rm cm}$ and $E_{||}\sim f_r\Phi_a/\rho_c$ for the curvature radius and the accelerating electric field, respectively, as illustrated in Figure~\ref{pratio}. The approximate death line is expressed as

\begin{equation}
  B_{*,d}\sim 6.2 (3+2\beta)^{-4/7} \times 10^8~{\rm G}
  \left(\frac{f_r}{0.2}\right)^{-3/7}
  \left(\frac{\rho_c}{10^9~{\rm cm}}\right)^{5/7}
  \left(\frac{P}{1~{\rm s}}\right)^{6/7}
  \left(\frac{R_{\rm WD}}{5\cdot10^8~{\rm cm}}\right)^{-13/7}.
\label{appen-death2}
\end{equation}
Since the curvature radius is expected to scale with
the stellar radius, $\rho_c\propto R_{\rm WD}$,
the death-line field scales approximately as
$B_{*,d}\propto R_{\rm WD}^{-8/7}$. To compare the death line with the acceleration region, Figure~\ref{fig:nondipole_gap} shows how $\ell_{\rm acc}+\ell_{\rm m}$ depends on $P$ and $B_{*,d}$, and the dashed line indicates the analytical death line given by equation~(\ref{appen-death2}) with $f_r=0.2$ and $\beta=5$. 

Finally, we evaluate the boundary above which synchrotron photons from secondary pairs can undergo magnetic pair creation. Near this boundary, the MFP of synchrotron photons becomes comparable to the stellar radius. Using equations~(\ref{mfp}) and~(\ref{eq:esyn}), we obtain the approximate boundary conditions
\begin{equation}
B \simeq 2.4\times10^9
\left( \frac{R_{\rm WD}}{5\cdot10^8~{\rm cm}}\right)^{-3/2}
\left(\frac{P}{1\,{\rm s}}\right)^{17/14}
\ {\rm G},
\label{B_syn1}
\end{equation}
for the dipole-field configuration, and

\begin{equation}
B \simeq 3.9\,(3+2\beta)^{-4/7}\times10^{9}
\left(\frac{f_r}{0.2}\right)^{-3/7}
\left(\frac{\rho_c}{10^9\,{\rm cm}}\right)^{5/7}
\left(\frac{P}{1\,{\rm s}}\right)^{6/7}
\left( \frac{R_{\rm WD}}{5\cdot10^8~{\rm cm}}\right)^{-13/7}
\ {\rm G},
\label{B_syn2}
\end{equation}
for the dipole--quadrupole configuration. 

\bibliography{sample631}{}

@ARTICLE{2026arXiv260604232R,
       author = {{Rose}, Kovi and {Pritchard}, Joshua and {Murphy}, Tara and {Driessen}, L.~N. and {Kaplan}, D.~L. and {Caleb}, M. and {Wang}, Ziteng and {Zic}, A. and {Andreoni}, I. and {Carney}, J. and {Barlow}, B.~N. and {Dobie}, D. and {Gu}, M. and {Heald}, G. and {Huber}, D. and {Lenc}, E. and {Leung}, J.~K. and {Lu}, W. and {Momose}, R. and {Pedersen}, M.~G. and {Qu}, Y. and {Rea}, N. and {de Ruiter}, I. and {Shaji}, K. and {Sivakoff}, G.~R. and {Thomson}, A.~J.~M. and {Wang}, Y.~L. and {Yang}, G.~J. and {Zahedy}, F.},
        title = "{Periodic Radio and X-ray Emission from an Accreting White Dwarf Binary}",
      journal = {arXiv e-prints},
         year = 2026,
        month = jun,
          eid = {arXiv:2606.04232},
        pages = {arXiv:2606.04232},
          doi = {10.48550/arXiv.2606.04232},
archivePrefix = {arXiv},
       eprint = {2606.04232},
 primaryClass = {astro-ph.HE},
       adsurl = {https://ui.adsabs.harvard.edu/abs/2026arXiv260604232R}
}

@ARTICLE{2024A&A...691A.179P,
       author = {{Pakmor}, R{\"u}diger and {Pelisoli}, Ingrid and {Justham}, Stephen and {Rajamuthukumar}, Abinaya S. and {R{\"o}pke}, Friedrich K. and {Schneider}, Fabian R.~N. and {de Mink}, Selma E. and {Ohlmann}, Sebastian T. and {Podsiadlowski}, Philipp and {Mor{\'a}n-Fraile}, Javier and {Vetter}, Marco and {Andrassy}, Robert},
        title = "{Large-scale ordered magnetic fields generated in mergers of helium white dwarfs}",
      journal = {\aap},
         year = 2024,
        month = nov,
       volume = {691},
          eid = {A179},
        pages = {A179},
          doi = {10.1051/0004-6361/202451352},
archivePrefix = {arXiv},
       eprint = {2407.02566},
 primaryClass = {astro-ph.SR},
       adsurl = {https://ui.adsabs.harvard.edu/abs/2024A&A...691A.179P}
}

@ARTICLE{2015ApJ...806L...1Z,
       author = {{Zhu}, Chenchong and {Pakmor}, R{\"u}diger and {van Kerkwijk}, Marten H. and {Chang}, Philip},
        title = "{Magnetized Moving Mesh Merger of a Carbon-Oxygen White Dwarf Binary}",
      journal = {\apjl},
         year = 2015,
        month = jun,
       volume = {806},
       number = {1},
          eid = {L1},
        pages = {L1},
          doi = {10.1088/2041-8205/806/1/L1},
archivePrefix = {arXiv},
       eprint = {1504.01732},
 primaryClass = {astro-ph.SR},
       adsurl = {https://ui.adsabs.harvard.edu/abs/2015ApJ...806L...1Z}
}

@ARTICLE{2013ApJ...773..136J,
       author = {{Ji}, Suoqing and {Fisher}, Robert T. and {Garc{\'\i}a-Berro}, Enrique and {Tzeferacos}, Petros and {Jordan}, George and {Lee}, Dongwook and {Lor{\'e}n-Aguilar}, Pablo and {Cremer}, Pascal and {Behrends}, Jan},
        title = "{The Post-merger Magnetized Evolution of White Dwarf Binaries: The Double-degenerate Channel of Sub-Chandrasekhar Type Ia Supernovae and the Formation of Magnetized White Dwarfs}",
      journal = {\apj},
         year = 2013,
        month = aug,
       volume = {773},
       number = {2},
          eid = {136},
        pages = {136},
          doi = {10.1088/0004-637X/773/2/136},
archivePrefix = {arXiv},
       eprint = {1302.5700},
 primaryClass = {astro-ph.SR},
       adsurl = {https://ui.adsabs.harvard.edu/abs/2013ApJ...773..136J}
}

@ARTICLE{2025arXiv250903216D,
       author = {{Desai}, Aayush and {Caiazzo}, Ilaria and {Vennes}, Stephane and {Kawka}, Adela and {Cunningham}, Tim and {Kotiwale}, Gauri and {Cristea}, Andrei A. and {Raymond}, John C. and {Camisassa}, Maria and {Althaus}, Leandro G. and {Hermes}, J.~J. and {Traulsen}, Iris and {Fuller}, James and {Heyl}, Jeremy and {van Roestel}, Jan and {Burdge}, Kevin B. and {Rodriguez}, Antonio C. and {Pelisoli}, Ingrid and {G{\"a}nsicke}, Boris T. and {Szkody}, Paula and {Maheshwari}, Sumit K. and {Vanderbosch}, Zachary P. and {Drake}, Andrew and {Ferrario}, Lilia and {Wickramasinghe}, Dayal and {Justham}, Stephen and {Pakmor}, Ruediger and {El-Badry}, Kareem and {Prince}, Thomas and {Kulkarni}, S.~R. and {Graham}, Matthew J. and {Rusholme}, Ben and {Laher}, Russ R. and {Purdum}, Josiah},
        title = "{Magnetic Atmospheres and Circumstellar Interaction in J1901+1458: Revisiting the Most Compact White Dwarf Merger Remnant in the light of new UV and X-ray data}",
      journal = {arXiv e-prints},
         year = 2025,
        month = sep,
          eid = {arXiv:2509.03216},
        pages = {arXiv:2509.03216},
          doi = {10.48550/arXiv.2509.03216},
archivePrefix = {arXiv},
       eprint = {2509.03216},
 primaryClass = {astro-ph.SR},
       adsurl = {https://ui.adsabs.harvard.edu/abs/2025arXiv250903216D}
}

@ARTICLE{2026A&A...706A.188C,
       author = {{Cristea}, Andrei A. and {Caiazzo}, Ilaria and {Cunningham}, Tim and {Raymond}, John C. and {Vennes}, Stephane and {Kawka}, Adela and {Desai}, Aayush and {Miller}, David R. and {Hermes}, J.~J. and {Fuller}, Jim and {Heyl}, Jeremy and {van Roestel}, Jan and {Burdge}, Kevin B. and {Rodriguez}, Antonio C. and {Pelisoli}, Ingrid and {G{\"a}nsicke}, Boris T. and {Szkody}, Paula and {Kenyon}, Scott J. and {Vanderbosch}, Zach and {Drake}, Andrew and {Ferrario}, Lilia and {Wickramasinghe}, Dayal and {Karambelkar}, Viraj R. and {Justham}, Stephen and {Pakmor}, Ruediger and {El-Badry}, Kareem and {Prince}, Thomas and {Kulkarni}, S.~R. and {Graham}, Matthew J. and {Masci}, Frank J. and {Groom}, Steven L. and {Purdum}, Josiah and {Dekany}, Richard and {Bellm}, Eric C.},
        title = "{A half ring of ionized circumstellar material trapped in the magnetosphere of a white dwarf merger remnant: A new class of white dwarf merger remnants with X-ray emission}",
      journal = {\aap},
         year = 2026,
        month = feb,
       volume = {706},
          eid = {A188},
        pages = {A188},
          doi = {10.1051/0004-6361/202556432},
archivePrefix = {arXiv},
       eprint = {2507.13850},
 primaryClass = {astro-ph.SR},
       adsurl = {https://ui.adsabs.harvard.edu/abs/2026A&A...706A.188C}
}

@ARTICLE{2012JPhG...39f5001B,
       author = {{Bednarek}, W.},
        title = "{Gamma-rays from electrons accelerated by rotating magnetized white dwarfs in globular clusters}",
      journal = {Journal of Physics G Nuclear Physics},
         year = 2012,
        month = jun,
       volume = {39},
       number = {6},
          eid = {065001},
        pages = {065001},
          doi = {10.1088/0954-3899/39/6/065001},
archivePrefix = {arXiv},
       eprint = {1203.4383},
 primaryClass = {astro-ph.HE},
       adsurl = {https://ui.adsabs.harvard.edu/abs/2012JPhG...39f5001B}
}

@ARTICLE{2025MNRAS.539.3013C,
       author = {{Cheng}, Yanchang and {Takata}, Jumpei},
        title = "{Spin evolution modelling for a newly formed white dwarf resulting from binary white dwarf merger}",
      journal = {\mnras},
         year = 2025,
        month = jun,
       volume = {539},
       number = {4},
        pages = {3013-3026},
          doi = {10.1093/mnras/staf580},
archivePrefix = {arXiv},
       eprint = {2504.03199},
 primaryClass = {astro-ph.HE},
       adsurl = {https://ui.adsabs.harvard.edu/abs/2025MNRAS.539.3013C}
}

@ARTICLE{2003ApJ...593.1040V,
       author = {{Vennes}, St{\'e}phane and {Schmidt}, G.~D. and {Ferrario}, L. and {Christian}, D.~J. and {Wickramasinghe}, D.~T. and {Kawka}, A.},
        title = "{A Multiwavelength Study of the High-Field Magnetic White Dwarf EUVE J0317-85.5 (=RE J0317-853)}",
      journal = {\apj},
         year = 2003,
        month = aug,
       volume = {593},
       number = {2},
        pages = {1040-1048},
          doi = {10.1086/376728},
       adsurl = {https://ui.adsabs.harvard.edu/abs/2003ApJ...593.1040V}
}

@ARTICLE{1984MNRAS.206..407M,
       author = {{Martin}, B. and {Wickramasinghe}, D.~T.},
        title = "{Magnetic field distributions in white dwarfs.}",
      journal = {\mnras},
         year = 1984,
        month = jan,
       volume = {206},
        pages = {407-422},
          doi = {10.1093/mnras/206.2.407},
       adsurl = {https://ui.adsabs.harvard.edu/abs/1984MNRAS.206..407M}
}

@ARTICLE{2005A&A...442..651E,
       author = {{Euchner}, F. and {Reinsch}, K. and {Jordan}, S. and {Beuermann}, K. and {G{\"a}nsicke}, B.~T.},
        title = "{Zeeman tomography of magnetic white dwarfs. II. The quadrupole-dominated magnetic field of <ASTROBJ>HE 1045-0908</ASTROBJ>}",
      journal = {\aap},
         year = 2005,
        month = nov,
       volume = {442},
       number = {2},
        pages = {651-660},
          doi = {10.1051/0004-6361:20053038},
archivePrefix = {arXiv},
       eprint = {astro-ph/0507631},
 primaryClass = {astro-ph},
       adsurl = {https://ui.adsabs.harvard.edu/abs/2005A&A...442..651E}
}

@ARTICLE{2024ApJ...974L..32C,
       author = {{Chernoglazov}, Alexander and {Philippov}, Alexander and {Timokhin}, Andrey},
        title = "{Coherence of Multidimensional Pair Production Discharges in Polar Caps of Pulsars}",
      journal = {\apjl},
         year = 2024,
        month = oct,
       volume = {974},
       number = {2},
          eid = {L32},
        pages = {L32},
          doi = {10.3847/2041-8213/ad7e24},
archivePrefix = {arXiv},
       eprint = {2409.15409},
 primaryClass = {astro-ph.HE},
       adsurl = {https://ui.adsabs.harvard.edu/abs/2024ApJ...974L..32C}
}

@ARTICLE{1983ApJ...266..215A,
       author = {{Arons}, J.},
        title = "{Pair creation above pulsar polar caps : geometrical structure and energetics of slot gaps.}",
      journal = {\apj},
         year = 1983,
        month = mar,
       volume = {266},
        pages = {215-241},
          doi = {10.1086/160771},
       adsurl = {https://ui.adsabs.harvard.edu/abs/1983ApJ...266..215A}
}

@ARTICLE{2024PASJ...76..702B,
       author = {{Bamba}, Aya and {Terada}, Yukikatsu and {Kashiyama}, Kazumi and {Kisaka}, Shota and {Minami}, Takahiro and {Takahashi}, Tadayuki},
        title = "{On the X-ray efficiency of the white dwarf pulsar candidate ZTF J190132.9+145808.7}",
      journal = {\pasj},
         year = 2024,
        month = aug,
       volume = {76},
       number = {4},
        pages = {702-707},
          doi = {10.1093/pasj/psae041},
archivePrefix = {arXiv},
       eprint = {2404.14722},
 primaryClass = {astro-ph.HE},
       adsurl = {https://ui.adsabs.harvard.edu/abs/2024PASJ...76..702B}
}

@ARTICLE{2017ApJ...851..143T,
       author = {{Takata}, J. and {Yang}, H. and {Cheng}, K.~S.},
        title = "{A Model for AR Scorpii: Emission from Relativistic Electrons Trapped by Closed Magnetic Field Lines of Magnetic White Dwarfs}",
      journal = {\apj},
         year = 2017,
        month = dec,
       volume = {851},
       number = {2},
          eid = {143},
        pages = {143},
          doi = {10.3847/1538-4357/aa9b33},
archivePrefix = {arXiv},
       eprint = {1712.03488},
 primaryClass = {astro-ph.HE},
       adsurl = {https://ui.adsabs.harvard.edu/abs/2017ApJ...851..143T}
}

@ARTICLE{2011PhRvD..83b3002K,
       author = {{Kashiyama}, Kazumi and {Ioka}, Kunihito and {Kawanaka}, Norita},
        title = "{White dwarf pulsars as possible cosmic ray electron-positron factories}",
      journal = {\prd},
         year = 2011,
        month = jan,
       volume = {83},
       number = {2},
          eid = {023002},
        pages = {023002},
          doi = {10.1103/PhysRevD.83.023002},
archivePrefix = {arXiv},
       eprint = {1009.1141},
 primaryClass = {astro-ph.HE},
       adsurl = {https://ui.adsabs.harvard.edu/abs/2011PhRvD..83b3002K}
}

@ARTICLE{1988SvAL...14..258U,
       author = {{Usov}, V.~V.},
        title = "{Generation of Gamma-Rays by a Rotating Magnetic White Dwarf}",
      journal = {Soviet Astronomy Letters},
         year = 1988,
        month = mar,
       volume = {14},
        pages = {258},
       adsurl = {https://ui.adsabs.harvard.edu/abs/1988SvAL...14..258U}
}

@ARTICLE{2023NatAs...7..931P,
       author = {{Pelisoli}, Ingrid and {Marsh}, T.~R. and {Buckley}, David A.~H. and {Heywood}, I. and {Potter}, Stephen. B. and {Schwope}, Axel and {Brink}, Jaco and {Standke}, Annie and {Woudt}, P.~A. and {Parsons}, S.~G. and {Green}, M.~J. and {Kepler}, S.~O. and {Munday}, James and {Romero}, A.~D. and {Breedt}, E. and {Brown}, A.~J. and {Dhillon}, V.~S. and {Dyer}, M.~J. and {Kerry}, P. and {Littlefair}, S.~P. and {Sahman}, D.~I. and {Wild}, J.~F.},
        title = "{A 5.3-min-period pulsing white dwarf in a binary detected from radio to X-rays}",
      journal = {Nature Astronomy},
         year = 2023,
        month = aug,
       volume = {7},
        pages = {931-942},
          doi = {10.1038/s41550-023-01995-x},
archivePrefix = {arXiv},
       eprint = {2306.09272},
 primaryClass = {astro-ph.SR},
       adsurl = {https://ui.adsabs.harvard.edu/abs/2023NatAs...7..931P}
}

@ARTICLE{2025NatAs...9..393L,
       author = {{Lee}, Y.~W.~J. and {Caleb}, M. and {Murphy}, Tara and {Lenc}, E. and {Kaplan}, D.~L. and {Ferrario}, L. and {Wadiasingh}, Z. and {Anumarlapudi}, A. and {Hurley-Walker}, N. and {Karambelkar}, V. and {Ocker}, S.~K. and {McSweeney}, S. and {Qiu}, H. and {Rajwade}, K.~M. and {Zic}, A. and {Bannister}, K.~W. and {Bhat}, N.~D.~R. and {Deller}, A. and {Dobie}, D. and {Driessen}, L.~N. and {Gendreau}, K. and {Glowacki}, M. and {Gupta}, V. and {Jahns-Schindler}, J.~N. and {Jaini}, A. and {James}, C.~W. and {Kasliwal}, M.~M. and {Lower}, M.~E. and {Shannon}, R.~M. and {Uttarkar}, P.~A. and {Wang}, Y. and {Wang}, Z.},
        title = "{The emission of interpulses by a 6.45-h-period coherent radio transient}",
      journal = {Nature Astronomy},
         year = 2025,
        month = mar,
       volume = {9},
        pages = {393-405},
          doi = {10.1038/s41550-024-02452-z},
archivePrefix = {arXiv},
       eprint = {2501.09133},
 primaryClass = {astro-ph.HE},
       adsurl = {https://ui.adsabs.harvard.edu/abs/2025NatAs...9..393L}
}

@ARTICLE{2026arXiv260110393R,
       author = {{Rea}, Nanda and {Hurley-Walker}, Natasha and {Caleb}, Manisha},
        title = "{Long Period Transients (LPTs): a comprehensive review}",
      journal = {arXiv e-prints},
         year = 2026,
        month = jan,
          eid = {arXiv:2601.10393},
        pages = {arXiv:2601.10393},
          doi = {10.48550/arXiv.2601.10393},
archivePrefix = {arXiv},
       eprint = {2601.10393},
 primaryClass = {astro-ph.HE},
       adsurl = {https://ui.adsabs.harvard.edu/abs/2026arXiv260110393R}
}

@ARTICLE{2015ApJ...810..144T,
       author = {{Timokhin}, A.~N. and {Harding}, A.~K.},
        title = "{On the Polar Cap Cascade Pair Multiplicity of Young Pulsars}",
      journal = {\apj},
         year = 2015,
        month = sep,
       volume = {810},
       number = {2},
          eid = {144},
        pages = {144},
          doi = {10.1088/0004-637X/810/2/144},
archivePrefix = {arXiv},
       eprint = {1504.02194},
 primaryClass = {astro-ph.HE},
       adsurl = {https://ui.adsabs.harvard.edu/abs/2015ApJ...810..144T}
}

@ARTICLE{1966RvMP...38..626E,
       author = {{Erber}, Thomas},
        title = "{High-Energy Electromagnetic Conversion Processes in Intense Magnetic Fields}",
      journal = {Reviews of Modern Physics},
         year = 1966,
        month = oct,
       volume = {38},
       number = {4},
        pages = {626-659},
          doi = {10.1103/RevModPhys.38.626},
       adsurl = {https://ui.adsabs.harvard.edu/abs/1966RvMP...38..626E}
}

@ARTICLE{1975ApJ...196...51R,
       author = {{Ruderman}, M.~A. and {Sutherland}, P.~G.},
        title = "{Theory of pulsars: polar gaps, sparks, and coherent microwave radiation.}",
      journal = {\apj},
         year = 1975,
        month = feb,
       volume = {196},
        pages = {51-72},
          doi = {10.1086/153393},
       adsurl = {https://ui.adsabs.harvard.edu/abs/1975ApJ...196...51R}
}

@ARTICLE{1971ApJ...164..529S,
       author = {{Sturrock}, P.~A.},
        title = "{A Model of Pulsars}",
      journal = {\apj},
         year = 1971,
        month = mar,
       volume = {164},
        pages = {529},
          doi = {10.1086/150865},
       adsurl = {https://ui.adsabs.harvard.edu/abs/1971ApJ...164..529S}
}

@ARTICLE{2022Natur.601..526H,
       author = {{Hurley-Walker}, N. and {Zhang}, X. and {Bahramian}, A. and {McSweeney}, S.~J. and {O'Doherty}, T.~N. and {Hancock}, P.~J. and {Morgan}, J.~S. and {Anderson}, G.~E. and {Heald}, G.~H. and {Galvin}, T.~J.},
        title = "{A radio transient with unusually slow periodic emission}",
      journal = {\nat},
         year = 2022,
        month = jan,
       volume = {601},
       number = {7894},
        pages = {526-530},
          doi = {10.1038/s41586-021-04272-x},
       adsurl = {https://ui.adsabs.harvard.edu/abs/2022Natur.601..526H}
}

@ARTICLE{1993ApJ...402..264C,
       author = {{Chen}, Kaiyou and {Ruderman}, Malvin},
        title = "{Pulsar Death Lines and Death Valley}",
      journal = {\apj},
         year = 1993,
        month = jan,
       volume = {402},
        pages = {264},
          doi = {10.1086/172129},
       adsurl = {https://ui.adsabs.harvard.edu/abs/1993ApJ...402..264C}
}

@ARTICLE{1969ApJ...157..869G,
       author = {{Goldreich}, Peter and {Julian}, William H.},
        title = "{Pulsar Electrodynamics}",
      journal = {\apj},
         year = 1969,
        month = aug,
       volume = {157},
        pages = {869},
          doi = {10.1086/150119},
       adsurl = {https://ui.adsabs.harvard.edu/abs/1969ApJ...157..869G}
}

@ARTICLE{1997ApJ...485..735M,
       author = {{Muslimov}, A. and {Harding}, A.~K.},
        title = "{Toward the Quasi-Steady State Electrodynamics of a Neutron Star}",
      journal = {\apj},
         year = 1997,
        month = aug,
       volume = {485},
       number = {2},
        pages = {735-746},
          doi = {10.1086/304457},
       adsurl = {https://ui.adsabs.harvard.edu/abs/1997ApJ...485..735M}
}

@ARTICLE{2021Natur.595...39C,
       author = {{Caiazzo}, Ilaria and {Burdge}, Kevin B. and {Fuller}, James and {Heyl}, Jeremy and {Kulkarni}, S.~R. and {Prince}, Thomas A. and {Richer}, Harvey B. and {Schwab}, Josiah and {Andreoni}, Igor and {Bellm}, Eric C. and {Drake}, Andrew and {Duev}, Dmitry A. and {Graham}, Matthew J. and {Helou}, George and {Mahabal}, Ashish A. and {Masci}, Frank J. and {Smith}, Roger and {Soumagnac}, Maayane T.},
        title = "{A highly magnetized and rapidly rotating white dwarf as small as the Moon}",
      journal = {\nat},
         year = 2021,
        month = jun,
       volume = {595},
       number = {7865},
        pages = {39-42},
          doi = {10.1038/s41586-021-03615-y},
archivePrefix = {arXiv},
       eprint = {2107.08458},
 primaryClass = {astro-ph.SR},
       adsurl = {https://ui.adsabs.harvard.edu/abs/2021Natur.595...39C}
}

@ARTICLE{2016Natur.537..374M,
       author = {{Marsh}, T.~R. and {G{\"a}nsicke}, B.~T. and {H{\"u}mmerich}, S. and {Hambsch}, F. -J. and {Bernhard}, K. and {Lloyd}, C. and {Breedt}, E. and {Stanway}, E.~R. and {Steeghs}, D.~T. and {Parsons}, S.~G. and {Toloza}, O. and {Schreiber}, M.~R. and {Jonker}, P.~G. and {van Roestel}, J. and {Kupfer}, T. and {Pala}, A.~F. and {Dhillon}, V.~S. and {Hardy}, L.~K. and {Littlefair}, S.~P. and {Aungwerojwit}, A. and {Arjyotha}, S. and {Koester}, D. and {Bochinski}, J.~J. and {Haswell}, C.~A. and {Frank}, P. and {Wheatley}, P.~J.},
        title = "{A radio-pulsing white dwarf binary star}",
      journal = {\nat},
         year = 2016,
        month = sep,
       volume = {537},
       number = {7620},
        pages = {374-377},
          doi = {10.1038/nature18620},
archivePrefix = {arXiv},
       eprint = {1607.08265},
 primaryClass = {astro-ph.SR},
       adsurl = {https://ui.adsabs.harvard.edu/abs/2016Natur.537..374M}
}

@ARTICLE{2023Natur.619..487H,
       author = {{Hurley-Walker}, N. and {Rea}, N. and {McSweeney}, S.~J. and {Meyers}, B.~W. and {Lenc}, E. and {Heywood}, I. and {Hyman}, S.~D. and {Men}, Y.~P. and {Clarke}, T.~E. and {Coti Zelati}, F. and {Price}, D.~C. and {Horv{\'a}th}, C. and {Galvin}, T.~J. and {Anderson}, G.~E. and {Bahramian}, A. and {Barr}, E.~D. and {Bhat}, N.~D.~R. and {Caleb}, M. and {Dall'Ora}, M. and {de Martino}, D. and {Giacintucci}, S. and {Morgan}, J.~S. and {Rajwade}, K.~M. and {Stappers}, B. and {Williams}, A.},
        title = "{A long-period radio transient active for three decades}",
      journal = {\nat},
         year = 2023,
        month = jul,
       volume = {619},
       number = {7970},
        pages = {487-490},
          doi = {10.1038/s41586-023-06202-5},
       adsurl = {https://ui.adsabs.harvard.edu/abs/2023Natur.619..487H}
}

@INPROCEEDINGS{2002nsps.conf..240U,
       author = {{Usov}, V.~V.},
        title = "{Two-stream Instability in Pulsar Magnetospheres}",
    booktitle = {Neutron Stars, Pulsars, and Supernova Remnants},
         year = 2002,
       editor = {{Becker}, W. and {Lesch}, H. and {Tr{\"u}mper}, J.},
        month = jan,
        pages = {240},
          doi = {10.48550/arXiv.astro-ph/0204402},
archivePrefix = {arXiv},
       eprint = {astro-ph/0204402},
 primaryClass = {astro-ph},
       adsurl = {https://ui.adsabs.harvard.edu/abs/2002nsps.conf..240U}
}

@ARTICLE{2001ApJ...560..871H,
       author = {{Hibschman}, Johann A. and {Arons}, Jonathan},
        title = "{Pair Production Multiplicities in Rotation-powered Pulsars}",
      journal = {\apj},
         year = 2001,
        month = oct,
       volume = {560},
       number = {2},
        pages = {871-884},
          doi = {10.1086/323069},
archivePrefix = {arXiv},
       eprint = {astro-ph/0107209},
 primaryClass = {astro-ph},
       adsurl = {https://ui.adsabs.harvard.edu/abs/2001ApJ...560..871H}
}

@ARTICLE{2002ApJ...565..482G,
       author = {{Gonthier}, Peter L. and {Ouellette}, Michelle S. and {Berrier}, Joel and {O'Brien}, Shawn and {Harding}, Alice K.},
        title = "{Galactic Populations of Radio and Gamma-Ray Pulsars in the Polar Cap Model}",
      journal = {\apj},
         year = 2002,
        month = jan,
       volume = {565},
       number = {1},
        pages = {482-499},
          doi = {10.1086/324535},
archivePrefix = {arXiv},
       eprint = {astro-ph/0109494},
 primaryClass = {astro-ph},
       adsurl = {https://ui.adsabs.harvard.edu/abs/2002ApJ...565..482G}
}

@ARTICLE{1996A&A...310..135Z,
       author = {{Zhang}, B. and {Qiao}, G.~J.},
        title = "{A study on pulsar inner-gap sparking comparing inverse Compton scattering and curvature radiation processes.}",
      journal = {\aap},
         year = 1996,
        month = jun,
       volume = {310},
        pages = {135-142},
       adsurl = {https://ui.adsabs.harvard.edu/abs/1996A&A...310..135Z}
}

@ARTICLE{2022MNRAS.516.5084B,
       author = {{Beskin}, V.~S. and {Istomin}, A. Yu},
        title = "{Pulsar death line revisited - II. 'The death valley'}",
      journal = {\mnras},
         year = 2022,
        month = nov,
       volume = {516},
       number = {4},
        pages = {5084-5091},
          doi = {10.1093/mnras/stac2423},
archivePrefix = {arXiv},
       eprint = {2207.04723},
 primaryClass = {astro-ph.HE},
       adsurl = {https://ui.adsabs.harvard.edu/abs/2022MNRAS.516.5084B}
}

@ARTICLE{2022NatAs...6..828C,
       author = {{Caleb}, Manisha and {Heywood}, Ian and {Rajwade}, Kaustubh and {Malenta}, Mateusz and {Stappers}, Benjamin Willem and {Barr}, Ewan and {Chen}, Weiwei and {Morello}, Vincent and {Sanidas}, Sotiris and {van den Eijnden}, Jakob and {Kramer}, Michael and {Buckley}, David and {Brink}, Jaco and {Motta}, Sara Elisa and {Woudt}, Patrick and {Weltevrede}, Patrick and {Jankowski}, Fabian and {Surnis}, Mayuresh and {Buchner}, Sarah and {Bezuidenhout}, Mechiel Christiaan and {Driessen}, Laura Nicole and {Fender}, Rob},
        title = "{Discovery of a radio-emitting neutron star with an ultra-long spin period of 76 s}",
      journal = {Nature Astronomy},
         year = 2022,
        month = may,
       volume = {6},
        pages = {828-836},
          doi = {10.1038/s41550-022-01688-x},
archivePrefix = {arXiv},
       eprint = {2206.01346},
 primaryClass = {astro-ph.HE},
       adsurl = {https://ui.adsabs.harvard.edu/abs/2022NatAs...6..828C}
}

@ARTICLE{2018ApJ...866...54T,
       author = {{Tan}, C.~M. and {Bassa}, C.~G. and {Cooper}, S. and {Dijkema}, T.~J. and {Esposito}, P. and {Hessels}, J.~W.~T. and {Kondratiev}, V.~I. and {Kramer}, M. and {Michilli}, D. and {Sanidas}, S. and {Shimwell}, T.~W. and {Stappers}, B.~W. and {van Leeuwen}, J. and {Cognard}, I. and {Grie{\ss}meier}, J. -M. and {Karastergiou}, A. and {Keane}, E.~F. and {Sobey}, C. and {Weltevrede}, P.},
        title = "{LOFAR Discovery of a 23.5 s Radio Pulsar}",
      journal = {\apj},
         year = 2018,
        month = oct,
       volume = {866},
       number = {1},
          eid = {54},
        pages = {54},
          doi = {10.3847/1538-4357/aade88},
archivePrefix = {arXiv},
       eprint = {1809.00965},
 primaryClass = {astro-ph.HE},
       adsurl = {https://ui.adsabs.harvard.edu/abs/2018ApJ...866...54T}
}

@ARTICLE{2018NatAs...2..865K,
       author = {{Keane}, E.~F.},
        title = "{The future of fast radio burst science}",
      journal = {Nature Astronomy},
         year = 2018,
        month = oct,
       volume = {2},
        pages = {865-872},
          doi = {10.1038/s41550-018-0603-0},
archivePrefix = {arXiv},
       eprint = {1811.00899},
 primaryClass = {astro-ph.HE},
       adsurl = {https://ui.adsabs.harvard.edu/abs/2018NatAs...2..865K}
}

@ARTICLE{2002ApJ...568..862H,
       author = {{Harding}, Alice K. and {Muslimov}, Alexander G.},
        title = "{Pulsar Polar Cap Heating and Surface Thermal X-Ray Emission. II. Inverse Compton Radiation Pair Fronts}",
      journal = {\apj},
         year = 2002,
        month = apr,
       volume = {568},
       number = {2},
        pages = {862-877},
          doi = {10.1086/338985},
archivePrefix = {arXiv},
       eprint = {astro-ph/0112392},
 primaryClass = {astro-ph},
       adsurl = {https://ui.adsabs.harvard.edu/abs/2002ApJ...568..862H}
}

@ARTICLE{2022MNRAS.510.2572B,
       author = {{Beskin}, V.~S. and {Litvinov}, P.~E.},
        title = "{Pulsar death line revisited - I. Almost vacuum gap}",
      journal = {\mnras},
         year = 2022,
        month = feb,
       volume = {510},
       number = {2},
        pages = {2572-2582},
          doi = {10.1093/mnras/stab3575},
archivePrefix = {arXiv},
       eprint = {2201.02875},
 primaryClass = {astro-ph.HE},
       adsurl = {https://ui.adsabs.harvard.edu/abs/2022MNRAS.510.2572B}
}

@ARTICLE{2018ApJ...853..106T,
       author = {{Takata}, J. and {Hu}, C. -P. and {Lin}, L.~C.~C. and {Tam}, P.~H.~T. and {Pal}, P.~S. and {Hui}, C.~Y. and {Kong}, A.~K.~H. and {Cheng}, K.~S.},
        title = "{A Non-thermal Pulsed X-Ray Emission of AR Scorpii}",
      journal = {\apj},
         year = 2018,
        month = feb,
       volume = {853},
       number = {2},
          eid = {106},
        pages = {106},
          doi = {10.3847/1538-4357/aaa23d},
archivePrefix = {arXiv},
       eprint = {1712.06341},
 primaryClass = {astro-ph.HE},
       adsurl = {https://ui.adsabs.harvard.edu/abs/2018ApJ...853..106T}
}

@ARTICLE{1998A&A...338..521I,
       author = {{Ikhsanov}, Nazar R.},
        title = "{The pulsar-like white dwarf in AE Aquarii}",
      journal = {\aap},
         year = 1998,
        month = oct,
       volume = {338},
        pages = {521-526},
       adsurl = {https://ui.adsabs.harvard.edu/abs/1998A&A...338..521I}
}

@ARTICLE{1978ApJ...222..297S,
       author = {{Scharlemann}, E.~T. and {Arons}, J. and {Fawley}, W.~M.},
        title = "{Potential drops above pulsar polar caps: ultrarelativistic particle acceleration along the curved magnetic field.}",
      journal = {\apj},
         year = 1978,
        month = may,
       volume = {222},
        pages = {297-316},
          doi = {10.1086/156144},
       adsurl = {https://ui.adsabs.harvard.edu/abs/1978ApJ...222..297S}
}

@ARTICLE{2024arXiv240707480D,
       author = {{Dong}, Fengqiu Adam and {Clarke}, Tracy and {Curtin}, Alice P. and {Kumar}, Ajay and {Stairs}, Ingrid and {Chatterjee}, Shami and {Cook}, Amanda M. and {Fonseca}, Emmanuel and {Gaensler}, B.~M. and {Hessels}, Jason W.~T. and {Kaspi}, Victoria M. and {Lazda}, Mattias and {Masui}, Kiyoshi W. and {McKee}, James W. and {Meyers}, Bradley W. and {Pearlman}, Aaron B. and {Ransom}, Scott M. and {Scholz}, Paul and {Shin}, Kaitlyn and {Smith}, Kendrick M. and {Tan}, Chia Min},
        title = "{The discovery of a nearby 421\raisebox{-0.5ex}\textasciitildes transient with CHIME/FRB/Pulsar}",
      journal = {arXiv e-prints},
         year = 2024,
        month = jul,
          eid = {arXiv:2407.07480},
        pages = {arXiv:2407.07480},
          doi = {10.48550/arXiv.2407.07480},
archivePrefix = {arXiv},
       eprint = {2407.07480},
 primaryClass = {astro-ph.HE},
       adsurl = {https://ui.adsabs.harvard.edu/abs/2024arXiv240707480D}
}

@ARTICLE{2006RPPh...69.2631H,
       author = {{Harding}, Alice K. and {Lai}, Dong},
        title = "{Physics of strongly magnetized neutron stars}",
      journal = {Reports on Progress in Physics},
         year = 2006,
        month = sep,
       volume = {69},
       number = {9},
        pages = {2631-2708},
          doi = {10.1088/0034-4885/69/9/R03},
archivePrefix = {arXiv},
       eprint = {astro-ph/0606674},
 primaryClass = {astro-ph},
       adsurl = {https://ui.adsabs.harvard.edu/abs/2006RPPh...69.2631H}
}

@ARTICLE{2025NatAs...9..672D,
       author = {{de Ruiter}, I. and {Rajwade}, K.~M. and {Bassa}, C.~G. and {Rowlinson}, A. and {Wijers}, R.~A.~M.~J. and {Kilpatrick}, C.~D. and {Stefansson}, G. and {Callingham}, J.~R. and {Hessels}, J.~W.~T. and {Clarke}, T.~E. and {Peters}, W. and {Wijnands}, R.~A.~D. and {Shimwell}, T.~W. and {ter Veen}, S. and {Morello}, V. and {Zeimann}, G.~R. and {Mahadevan}, S.},
        title = "{Sporadic radio pulses from a white dwarf binary at the orbital period}",
      journal = {Nature Astronomy},
         year = 2025,
        month = may,
       volume = {9},
        pages = {672-684},
          doi = {10.1038/s41550-025-02491-0},
archivePrefix = {arXiv},
       eprint = {2408.11536},
 primaryClass = {astro-ph.HE},
       adsurl = {https://ui.adsabs.harvard.edu/abs/2025NatAs...9..672D}
}

@ARTICLE{2025A&A...695L...8R,
       author = {{Rodriguez}, Antonio C.},
        title = "{Spectroscopic detection of a 2.9-hour orbit in a long-period radio transient}",
      journal = {\aap},
         year = 2025,
        month = mar,
       volume = {695},
          eid = {L8},
        pages = {L8},
          doi = {10.1051/0004-6361/202553684},
archivePrefix = {arXiv},
       eprint = {2501.03315},
 primaryClass = {astro-ph.SR},
       adsurl = {https://ui.adsabs.harvard.edu/abs/2025A&A...695L...8R}
}

@ARTICLE{1982ApJ...254..713B,
       author = {{Barnard}, J.~J. and {Arons}, J.},
        title = "{Pair production and pulsar cutoff in magnetized neutron stars with nondipolar magnetic geometry}",
      journal = {\apj},
         year = 1982,
        month = mar,
       volume = {254},
        pages = {713-734},
          doi = {10.1086/159784},
       adsurl = {https://ui.adsabs.harvard.edu/abs/1982ApJ...254..713B}
}

@book{RybickiLightman1979,
  author = {Rybicki, George B. and Lightman, Alan P.},
  title = {Radiative Processes in Astrophysics},
  publisher = {Wiley},
  address = {New York},
  year = {1979}
}

@ARTICLE{2026JHEAp..5300593M,
       author = {{Malheiro}, Manuel and {Borges}, Sarah V. and {Coelho}, Jaziel G. and {Kianfar}, Khashayar and {Lobato}, Ronaldo V. and {Otoniel}, Edson and {Rueda}, Jorge A. and {Sousa}, Manoel F. and {Weber}, Fridolin},
        title = "{Double white dwarf mergers as progenitors of long-period transients}",
      journal = {Journal of High Energy Astrophysics},
         year = 2026,
        month = jul,
       volume = {53},
          eid = {100593},
        pages = {100593},
          doi = {10.1016/j.jheap.2026.100593},
archivePrefix = {arXiv},
       eprint = {2603.08416},
 primaryClass = {astro-ph.HE},
       adsurl = {https://ui.adsabs.harvard.edu/abs/2026JHEAp..5300593M}
}

@ARTICLE{2008ApJ...683..466V,
       author = {{Valyavin}, G. and {Wade}, G.~A. and {Bagnulo}, S. and {Szeifert}, T. and {Landstreet}, J.~D. and {Han}, Inwoo and {Burenkov}, A.},
        title = "{The Peculiar Magnetic Field Morphology of the White Dwarf WD 1953-011: Evidence for a Large-Scale Magnetic Flux Tube?}",
      journal = {\apj},
         year = 2008,
        month = aug,
       volume = {683},
       number = {1},
        pages = {466-478},
          doi = {10.1086/589234},
archivePrefix = {arXiv},
       eprint = {0804.3140},
 primaryClass = {astro-ph},
       adsurl = {https://ui.adsabs.harvard.edu/abs/2008ApJ...683..466V}
}

@ARTICLE{2023ApJ...958..134S,
       author = {{Sousa}, M.~F. and {Coelho}, J.~G. and {de Araujo}, J.~C.~N. and {Guidorzi}, C. and {Rueda}, J.~A.},
        title = "{On the Optical Transients from Double White-dwarf Mergers}",
      journal = {\apj},
         year = 2023,
        month = dec,
       volume = {958},
       number = {2},
          eid = {134},
        pages = {134},
          doi = {10.3847/1538-4357/ad022f},
archivePrefix = {arXiv},
       eprint = {2310.06655},
 primaryClass = {astro-ph.SR},
       adsurl = {https://ui.adsabs.harvard.edu/abs/2023ApJ...958..134S}
}

@article{GarciaBerro2012,
  author       = {Garc{\'i}a-Berro, E. and Lor{\'e}n-Aguilar, P. and Aznar-Sigu{\'a}n, G. and Torres, S. and Camacho, J. and Althaus, L. G. and C{\'o}rsico, A. H. and K{\"u}lebi, B. and Isern, J.},
  title        = {Double Degenerate Mergers as Progenitors of High-Field Magnetic White Dwarfs},
  journal      = {The Astrophysical Journal},
  volume       = {749},
  pages        = {25},
  year         = {2012},
  doi          = {10.1088/0004-637X/749/1/25},
  eprint       = {1201.2411},
  archivePrefix= {arXiv},
  primaryClass = {astro-ph.SR}
}

@article{Katz2022,
  author       = {Katz, J. I.},
  title        = {Highly Magnetized White Dwarfs},
  journal      = {Astrophysics and Space Science},
  volume       = {367},
  pages        = {108},
  year         = {2022},
  doi          = {10.1007/s10509-022-04174-2}
}

@ARTICLE{2020ApJ...895...26B,
       author = {{Borges}, Sarah V. and {Rodrigues}, Claudia V. and
                 {Coelho}, Jaziel G. and {Malheiro}, Manuel and
                 {Castro}, Manuel},
        title = "{A Magnetic White Dwarf Accretion Model for the Anomalous X-Ray Pulsar 4U 0142+61}",
      journal = {The Astrophysical Journal},
         year = 2020,
       volume = {895},
       number = {1},
        pages = {26},
          doi = {10.3847/1538-4357/ab8add}
}

@ARTICLE{2024ApJ...961..214R,
       author = {{Rea}, N. and {Hurley-Walker}, N. and {Pardo-Araujo}, C. and {Ronchi}, M. and {Graber}, V. and {Coti Zelati}, F. and {de Martino}, D. and {Bahramian}, A. and {McSweeney}, S.~J. and {Galvin}, T.~J. and {Hyman}, S.~D. and {Dall'Ora}, M.},
        title = "{Long-period Radio Pulsars: Population Study in the Neutron Star and White Dwarf Rotating Dipole Scenarios}",
      journal = {\apj},
         year = 2024,
        month = feb,
       volume = {961},
       number = {2},
          eid = {214},
        pages = {214},
          doi = {10.3847/1538-4357/ad165d},
archivePrefix = {arXiv},
       eprint = {2307.10351},
 primaryClass = {astro-ph.HE},
       adsurl = {https://ui.adsabs.harvard.edu/abs/2024ApJ...961..214R}
}
\bibliographystyle{aasjournal}

\end{document}